\documentclass[12pt]{article}
\usepackage{graphicx}
\usepackage{amsmath}
 
\begin{document}

\title{Outflow probability for drift--diffusion
dynamics} 

\author{Julia Hinkel, Reinhard Mahnke}

\maketitle

\centerline{Institute of Physics, Rostock University, D
-- 18051, Rostock, Germany}

\centerline{julia.hinkel@uni-rostock.de} 

\begin{abstract}
The proposed explanations are provided for the
one--dimensional diffusion process with constant drift by
using forward Fokker--Planck technique. We  present the
exact calculations and numerical evaluation to get the
outflow probability in a finite interval, i.~e. first
passage time probability density distribution taking into
account reflecting boundary on left hand side and absorbing
border on right hand side. This quantity is
calculated from balance equation which follows from
conservation of probability. At first, the
initial--boundary--value problem is solved analytically in
terms of eigenfunction expansion which relates to
Sturm--Liouville analysis. The results are obtained
for all possible values of drift (positive, zero,
negative). As application we get the cumulative
breakdown probability which is used in theory of traffic
flow.
\end{abstract}
\centerline{PACS numbers: 02.60.Lj, 02.50.−-r, 02.50.Fz,
02.50.Ga}
\section{Introduction} 
Nowadays, the natural sciences deal with objects which have
nondeterministic behaviour. Their descriptions can be found
in theory of stochastic processes as a branch of probability
theory~\cite{gard}. There are different
theoretical approaches
for similar investigations using language of
stochastic trajectories as well as probability
distributions. The fundamental equation which gives us 
the probabilistic description is the Fokker--Planck
equation~\cite{gard,ris}. Our motivations for
theoretical investigations in this field are given by
application of the models of many--particles system
which are considered in theoretical physics, i.~e. physics
of traffic flow~\cite{MKL}. Here we would like to find an
analytical solution for the special case when the stochastic
variable belongs to a finite interval in terms of
probability density distributions as well as cumulative
probability~\cite{red}. The interval is
defined as closed on left hand side and opened on right hand
side. Due to these properties we introduce boundary
conditions which determine the behaviour of the solution.
Another important analysed quantity is the
outflow (or breakdown) probability at right border which is
found from the solution of Fokker--Planck equation by using
balance equation. 

\begin{figure}
\begin{center}
\includegraphics[scale=0.35]{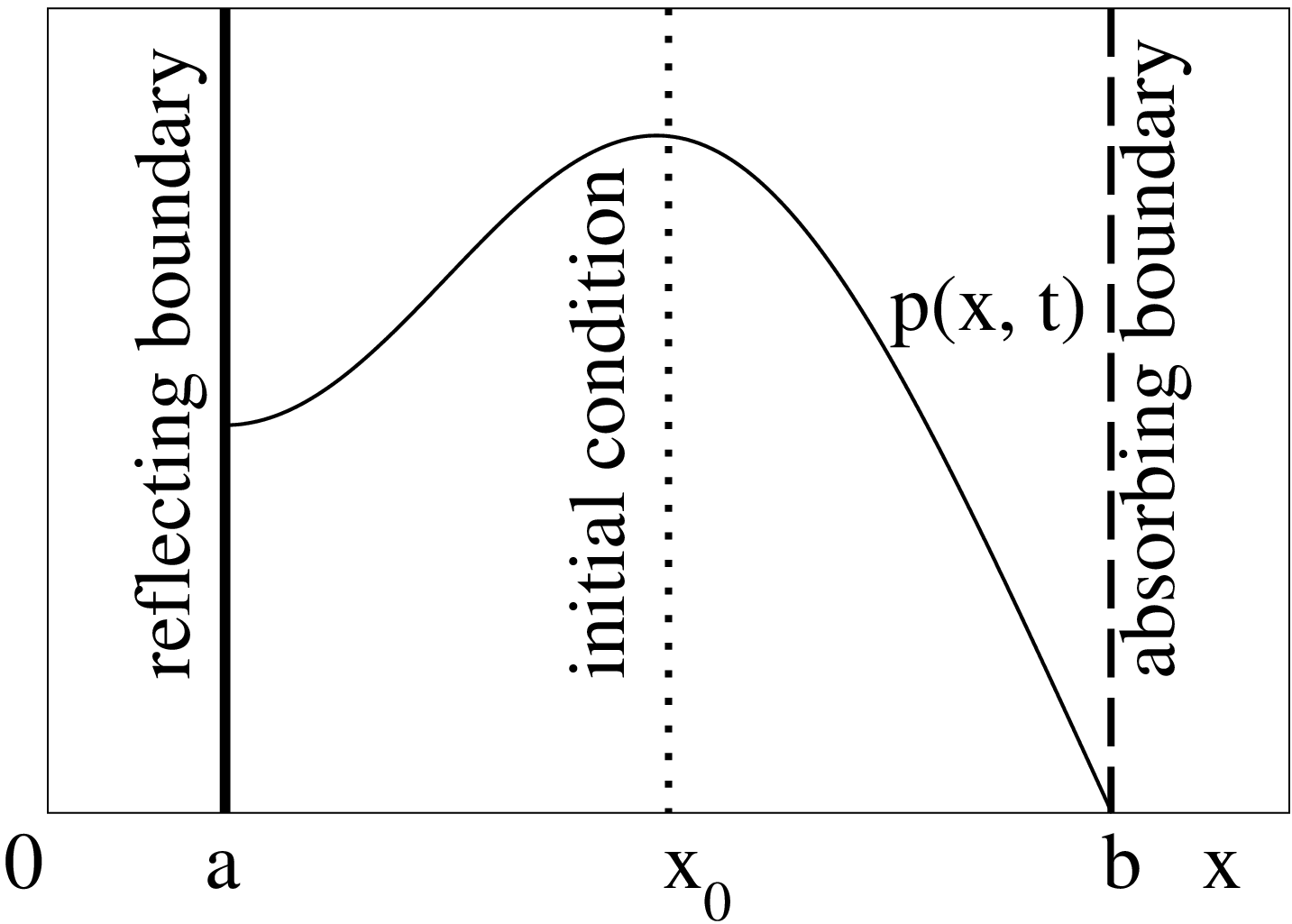}
\end{center}
\caption
{\label{schema}
Schematic picture of the boundary--value problem showing
the probability density $p(x\,,t)$ in the interval $a\le
x\le b$.}
\end{figure}

Let us consider the initial--boundary--value--problem (shown
schematically in Fig.~\ref{schema}) with constant
diffusion coefficient $D$ and constant drift coefficient
$v$.
Our task is to calculate the probability density $p(x,t)$ to
find the system
in state $x$ (exact in interval $[x;x+dx]$) at time moment
$t$.
The dynamics of $p(x,t)$ is given by the forward
drift--diffusion--equation as well as initial and boundary
conditions~\cite{gard} by following dynamics
\begin{eqnarray}
&&\frac{\partial p(x,t)}{\partial t} =
- v\,\frac{\partial p(x,t)}{\partial x}
+D\,\frac{\partial^2 p(x,t)}{\partial x^2} \; ,
\label{eq:ddeq}\\
&& \mbox{or} \quad \frac{\partial p(x,t)}{\partial t} +
\frac{\partial
j(x,t)}{\partial x} = 0 \\
 \mbox{with flux} &&\quad j(x,t)= \,v\, p(x,t) -
D\,\frac{\partial p(x,t)}{\partial x} \label{flux}
\end{eqnarray}
with the initial condition
\begin{equation}
p(x,t=0) = \delta(x-x_0) \; ,
\label{eq:init}
\end{equation}
and two boundary conditions~\cite{gard}, i.~e. reflecting
boundary at
$x=a$ 
\begin{equation}
j(x=a,t) = v\,p(x=a,t) -
D\, \left.\frac{\partial p(x,t)}{\partial x}\right|_{x=a} =
0
\; ,
\label{eq:reflect13}
\end{equation}
and absorbing boundary at $x=b$ 
\begin{equation}
p(x=b,t) = 0 \; .
\label{eq:absorb13}
\end{equation}

It is convenient to formulate the drift--diffusion problem
in dimensionless variables. For this purpose we
define new variables $y$ and $T$ by
\begin{eqnarray}
y = \frac{x-a}{b-a} &\quad \mbox{and}\quad & T =
\frac{D}{(b-a)^2}
\, t\,.
\end{eqnarray}
As a result, the system of partial
differential equations~(\ref{eq:ddeq})
 -- (\ref{eq:absorb13})
can be rewritten as 
\begin{equation}
\frac{\partial P(y,T)}{\partial T} =
- \Omega\, \frac{\partial P(y,T)}{\partial y}
+ \frac{\partial^2 P(y,T)}{\partial y^2} \; ,
\label{eq:dimldd}
\end{equation}
with initial condition 
\begin{equation}
P(y,T=0) = \delta(y-y_0) \; ,
\label{eq:25a}
\end{equation}
reflecting boundary at $y=0$
\begin{equation}
J(y=0,T) = \Omega \,P(y=0,T) -
 \left.\frac{\partial P(y,T)}{\partial y}\right|_{y=0} = 0
\; ,
\end{equation}
and absorbing boundary at $y=1$ 
\begin{equation}
P(y=1,T) = 0 \; .
\label{eq:dimlabs}
\end{equation}
Hence, our problem has only one dimensionless control
parameter $\Omega = \displaystyle{\frac{v}{D}} \left(b -
a\right)$ (scaled drift $v$ which
may have positive, zero, or negative values). The parameter
$\Omega$ has the same meaning as P\'eclet number which has
been used in \cite{red}. 

The system of equations (\ref{eq:dimldd}) --
(\ref{eq:dimlabs}) will be solved exactly by applying the
forward technique~\cite{gard}. The main idea is to obtain
the solution of Fokker--Planck equation and after that the
first passage time distribution in terms of
probability density. Both quantities will be presented
as eigenfunction expantions. The survival probability
and moments of first passage time can be calculated
differently by using backward drift--diffusion
equation. These results shown in~\cite{phr2,phr1,red}
do not give the complete solution of the problem under
consideration. Our presented analysis of the
reference system (\ref{eq:dimldd}) -- (\ref{eq:dimlabs}) is
the key result in order to study more complicated
situations with  nonlinear drift function $\Omega(y)$.
%%%%%%%%%%
\section{Solution in terms of orthogonal eigenfunctions}
To find the solution of the well--defined drift--diffusion
problem, first we take the
dimensionless form~(\ref{eq:dimldd}) -- (\ref{eq:dimlabs})
and
use a transformation to a new function $Q$ by
\begin{equation}
Q(y,T) = e^{-\frac{\Omega}{2} y} \,P(y,T) \;.
\label{eq:subst}
\end{equation}
This results in a dynamics without first derivative called
reduced Fokker--Planck--equation
\begin{equation}
\frac{\partial Q(y,T)}{\partial T} =
- \frac{\Omega^2}{4}\, Q(y,T)
+ \frac{\partial^2 Q(y,T)}{\partial y^2} \; .
\label{eq:13.17}
\end{equation}
According to~(\ref{eq:subst}) the initial condition is
transformed to
\begin{equation}
Q(y,T=0) = e^{-\frac{\Omega}{2} y_0}\, P(y,T=0) \;,
\end{equation}
whereas the reflecting boundary condition at $y=0$
becomes
\begin{equation}
\frac{\Omega}{2} \,Q(y=0,T) -
 \left.\frac{\partial Q(y,T)}{\partial y}\right|_{y=0} = 0
\;,
\label{eq:refbc}
\end{equation}
and the absorbing boundary condition at $y=1$ now reads
\begin{equation}
Q(y=1,T) = 0 \;.
\label{eq:absbc}
\end{equation}

The solution of reduced equation (\ref{eq:13.17}) can be
found
by the method of separation of variables~\cite{pde}.
Making a separation ansatz $Q(y,T) = \chi(T) \psi(y)$, we
obtain
\begin{equation} 
\frac{1}{\chi(T)} \frac{d \chi(T)}{dT} =
- \frac{\Omega^2}{4} + \frac{1}{\psi(y)} \frac{d^2
\psi(y)}{dy^2} \; .
\label{eq:13.34}
\end{equation}
Both sides should be equal to a constant. This constant is
denoted by $-
\lambda$, where $\lambda$ has the meaning of an eigenvalue.
The
eigenvalue $\lambda$ should be real and nonnegative.

Integration of the left hand side gives exponential decay
\begin{equation}
\chi(T) = \chi_0 \exp\{- \lambda \, T\}
\end{equation}
with $\chi(T=0) = \chi_0$ and setting $\chi_0 = 1$.

Let us now define the dimensionless wave number $k$ as $k^2
= \lambda$.
The right--hand side of eq.~(\ref{eq:13.34}) then transforms
into the
following wave equation
\begin{equation}
\frac{d^2 \psi(y)}{d y^2} +
\left( k^2 - \frac{\Omega^2}{4} \right) \psi(y) = 0 \;.
\label{eq:wave}
\end{equation}
Further on, we introduce a modified wave number
$\tilde{k}^2 = k^2 - \Omega^2/4$.
Note that $\tilde{k} = + \sqrt{k^2 - \Omega^2/4}$
may be complex (either pure real or pure imaginary).

First we consider the case where $\tilde{k}$ is real.
A suitable complex ansatz for the solution of the wave
equation~(\ref{eq:wave}) reads
\begin{equation}
\psi(y) = C^* \exp\{+ i \tilde{k} y \} +
C \exp\{- i \tilde{k} y \}
\label{eq:form}
\end{equation}
with complex coefficients $C=A/2 + i \, B/2$ and $C^*=A/2 -
i \, B/2$
chosen in such a way to ensure a real solution
\begin{equation}
\psi(y) = A \cos(\tilde{k} y) + B \sin(\tilde{k} y) \;.
\label{eq:sol}
\end{equation}

The two boundary conditions~(\ref{eq:refbc})
and~(\ref{eq:absbc})
can be used to determine the modified wave
number $\tilde{k}$ and the ratio $A/B$. The particular
solutions are
eigenfunctions $\psi_m(y)$, which form a complete set of
orthogonal
functions. As the third condition, we require that these
eigenfunctions
are normalised
\begin{equation}
\int\limits_0^1 \psi_m^2(y) dy = 1 \; .
\label{eq:norm}
\end{equation}
In this case all three parameters $\tilde k$, $A$, and
$B$ are defined.

The condition for the left boundary (\ref{eq:refbc}) reads
\begin{equation} \label{eq:13.40}
\frac{\Omega}{2} \psi(y=0) -
\left.\frac{d \psi(y)}{d y}\right|_{y=0} = 0 \;.
\end{equation}
After a substitution by~(\ref{eq:form}) it reduces to
\begin{equation}
\frac{\Omega}{2} (C^* + C) = i \tilde{k} \, (C^* - C)
\end{equation}
or
\begin{equation}
\frac{\Omega}{2} A = \tilde{k} B \;.
\label{eq:1}
\end{equation}

The condition for the right boundary (\ref{eq:absbc})
\begin{equation} \label{eq:13.43}
\psi(y=1) = 0
\end{equation}
gives us
\begin{equation}
C^* \exp\{+ i \tilde{k} \} +
C \exp\{- i \tilde{k} \}  = 0
\end{equation}
or
\begin{equation}
A \cos\left(\tilde{k} \right)
+ B \sin\left(\tilde{k} \right) = 0 \;.
\label{eq:2}
\end{equation}

By putting both equalities~(\ref{eq:1}) and~(\ref{eq:2})
together
and looking for a nontrivial solution,
we arrive at a transcendental equation
\begin{equation}
i \frac{\Omega}{2}
\left( \exp\{+ i \tilde{k} \} - \exp\{- i \tilde{k} \}
\right) =
\tilde{k} \left( \exp\{+ i \tilde{k} \} + \exp\{- i
\tilde{k} \} \right)
\end{equation}
or
\begin{equation}
\frac{\Omega}{2} \sin\left(\tilde{k} \right)
+ \tilde{k}  \cos\left(\tilde{k} \right) = 0 \; ,
\label{eq:transc}
\end{equation}
respectively
\begin{equation}
\tan \left( \tilde{k} \right) = - \frac{2}{\Omega} \,
\tilde{k} \; ,
\label{eq:13-48}
\end{equation}
which gives the spectrum of values $\tilde{k}_m$ with $m=0,
1, 2, \ldots$
(numbered in such a way that
$0<\tilde k_0 < \tilde k_1 < \tilde k_2 < \ldots$)
and the discrete eigenvalues $\lambda_m > 0\,$.

Due to~(\ref{eq:sol}) and (\ref{eq:2}), the eigenfunctions
can be written as
\begin{equation}
\psi_m(y)= R_m \left[ \cos \left(\tilde k_m y \right) \sin
\left(\tilde k_m
\right) - \cos \left(\tilde k_m \right) \sin \left(\tilde
k_m y \right) \right] \; ,
\label{eq:Hmm}
\end{equation}
where $R_m= A_m/ \sin \left(\tilde k_m \right)=
-B_m/ \cos \left(\tilde k_m \right)$.
Taking into account the identity $\sin(\alpha-\beta)=\sin
\alpha \cos \beta -
\cos \alpha \sin \beta$, eq.~(\ref{eq:Hmm}) reduces to
\begin{equation}
\psi_m(y)= R_m \sin \left[ \tilde k_m (1-y) \right] \; .
\label{eq:H1}
\end{equation}
The normalisation constant $R_m$ is found by
inserting~(\ref{eq:H1})
into~(\ref{eq:norm}).  Calculation of the normalisation
integral by
using the transcendental equation~(\ref{eq:transc}) gives
us
\begin{eqnarray}
&&R_m^2 \int_0^{1} \sin^2\left[\tilde{k}_m \left(1-y\right)
\right] dy
= R_m^2 \left[ \frac{1}{2} - \frac{1}{4 \tilde k_m} \sin
\left( 2 \tilde
k_m \right) \right] \nonumber \\
&&= \frac{R_m^2}{2} \left(1+\frac{\Omega}{2}
\frac{1}{\tilde{k}_m^2 + \Omega^2/4}\right) = 1 \;,
\label{eq:Rm}
\end{eqnarray}
and hence (\ref{eq:H1}) becomes
\begin{equation}
\psi_m(y)= \sqrt{\frac{2}{1+ \frac{\Omega}{2}
\frac{1}{\tilde{k}_m^2+\Omega^2/4}} }
\sin \left[ \tilde k_m (1-y) \right]
\label{eq:Hn}
\end{equation}
or
\begin{equation}
\psi_m(y)= \sqrt{\frac{2}{1+ \frac{\Omega}{2}
\frac{1}{k_m^2} }}
\sin \left[ \sqrt{k_m^2-\Omega^2/4} \; (1-y) \right] \;.
\end{equation}
This calculation refers to the case $\Omega > -2$ where all
wave numbers $k_m$ or $\tilde k_m= \sqrt{k_m^2 -
\Omega^2/4}$
are real and positive.

\begin{figure}
\begin{center}
\includegraphics[scale=0.29]{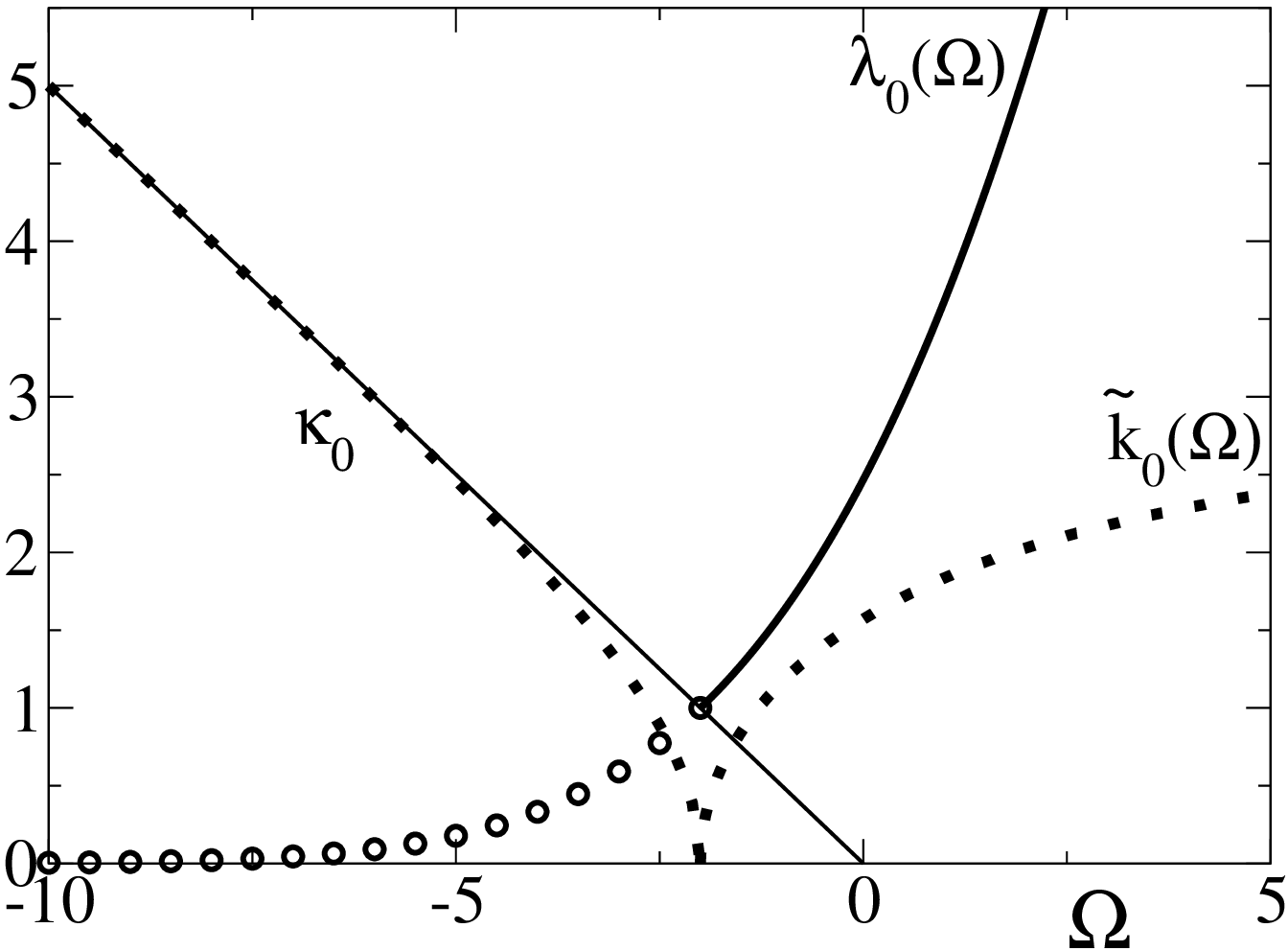}
\end{center}
\caption
{\label{grund}
The wave number ${\tilde k}_0$ ($\Omega \ge-\,2$)
respectively
$\kappa_0$ ($\Omega \le -\,2$) and eigenvalue $\lambda_0$
for ground
state $m = 0$. The thin straight line shows the
approximation
$\kappa_0 \approx -\,\Omega/2$ valid for large negative
$\Omega < -\,5$.}
\end{figure}

However the smallest or ground--state wave vector $\tilde
k_0$
vanishes when $\Omega$ tends to $-2$ from above,
and no continuation of this solution exists
on the real axis for $\Omega < -2$. A purely imaginary
solution
$\tilde k_0 = i \kappa_0$ appears instead, where $\kappa_0$
is real, see
Fig.~\ref{grund}.
In this case
(for $\Omega < -2$) a real ground--state eigenfunction
$\psi_0(y)$ can be
found in the form~(\ref{eq:form}) where $C=A/2 + B/2$ and
$C^*=A/2-B/2$, i.~e.,
\begin{equation}
\psi_0(y) = A \cosh(\kappa_0 y) + B \sinh(\kappa_0 y) \; .
\end{equation}
The transcendental equation for the wave number $\tilde k_0=
i \kappa_0$ can be
written as the following equation for $\kappa_0$
\begin{equation}
\frac{\Omega}{2} \sinh \left( \kappa_0 \right) + 
\kappa_0 \cosh \left( \kappa_0 \right) = 0 \; .
\label{eq:13-55}
\end{equation}

\begin{figure}
\begin{center}
\hspace*{\fill}
\includegraphics[scale=0.25]{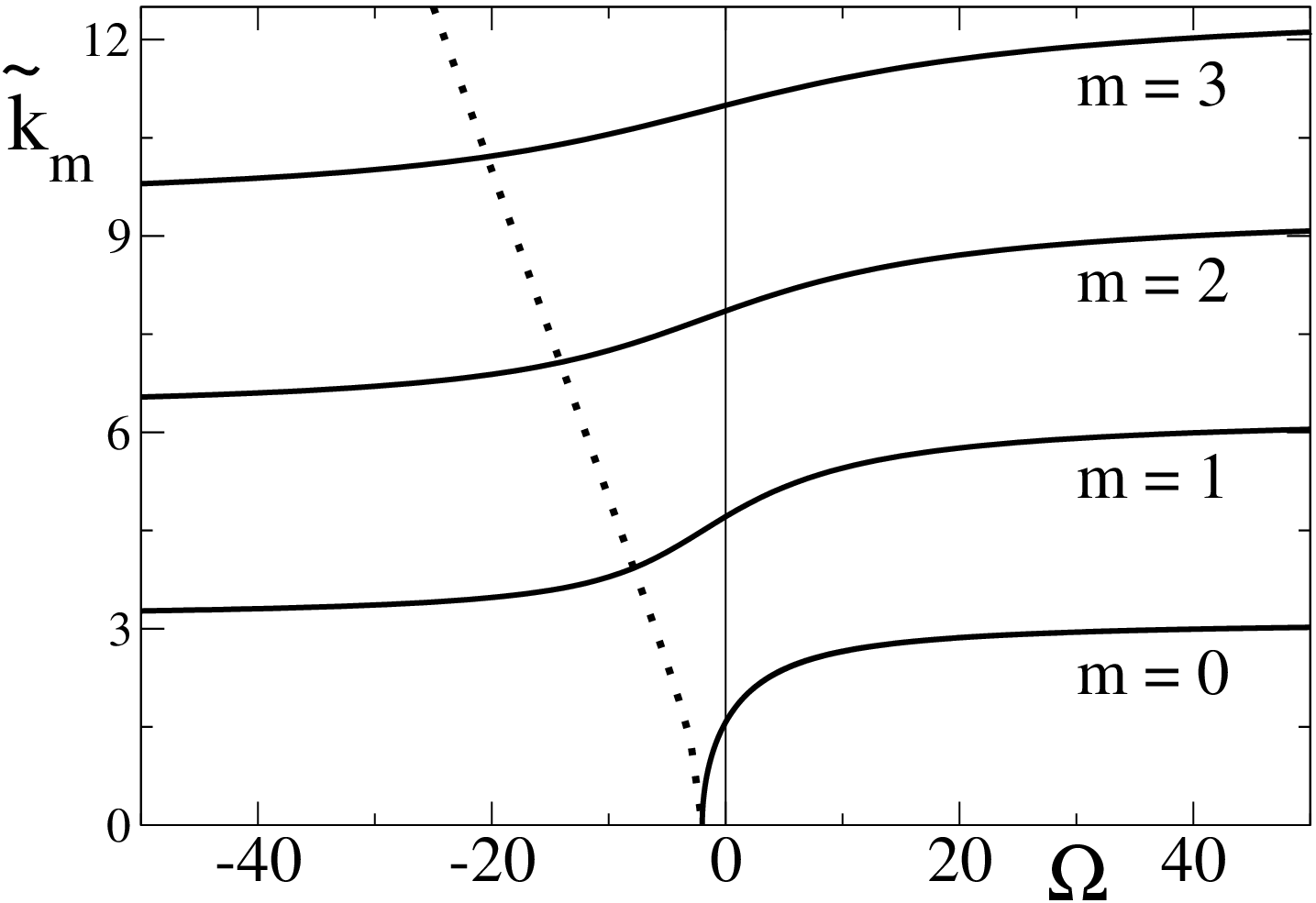}
\hspace*{\fill}
\includegraphics[scale=0.25]{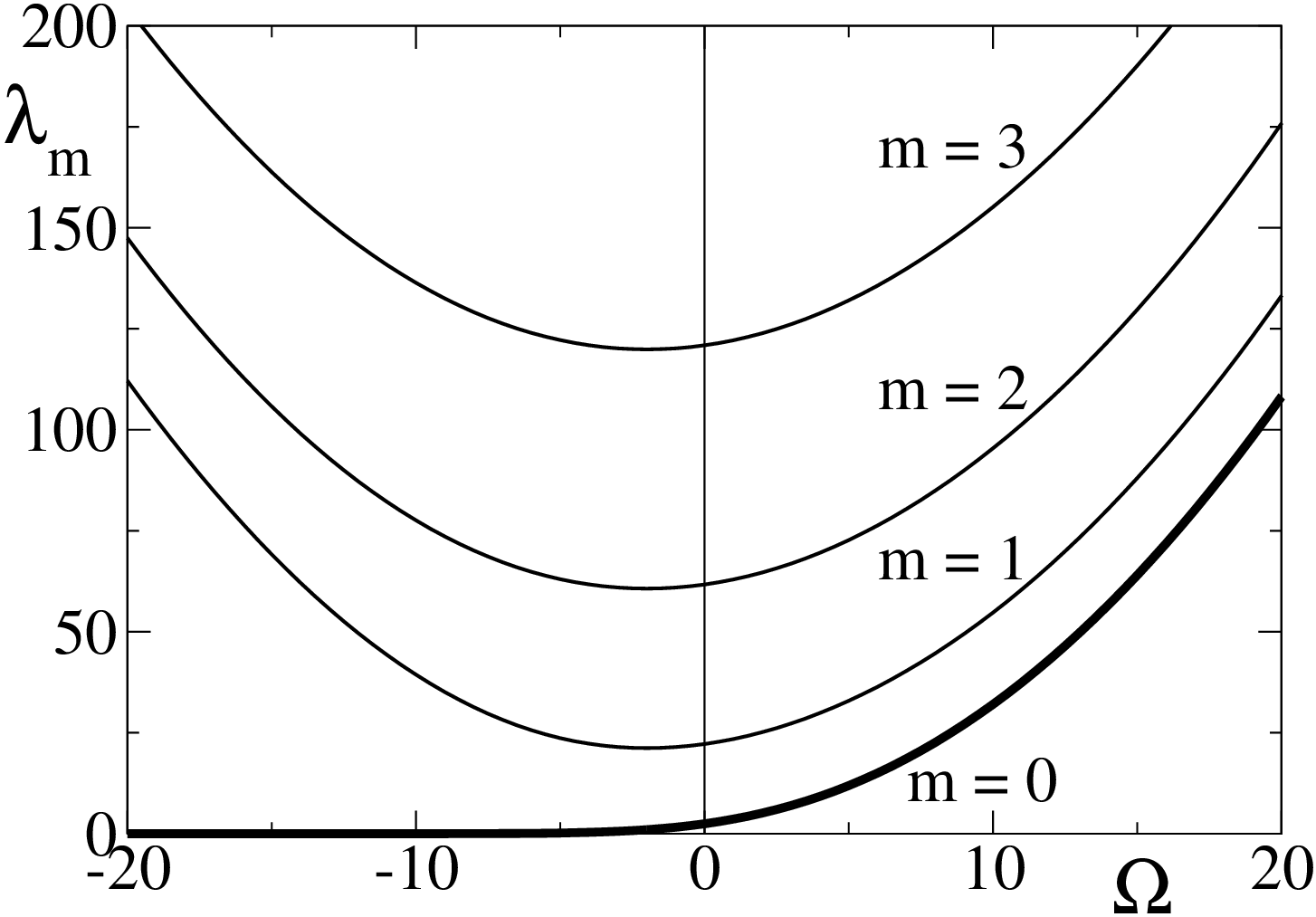}
\hspace*{\fill}
\end{center}
\caption
{\label{kandlambda}
The parameter dependence of wave numbers ${\tilde
k}_m(\Omega)$ and
eigenvalues $\lambda_m(\Omega)$ for ground state $m = 0$ and
excited states $m = 1, 2, 3$.}
\end{figure}

As compared to the previous case $\Omega >-2$,
trigonometric functions are replaced by the corresponding
hyperbolic
ones. Similar calculations as before yield
\begin{equation}
\psi_0(y)= \sqrt{-\frac{2}{1+ \frac{\Omega}{2}
\frac{1}{-\kappa_0^2+\Omega^2/4}} }
\sinh \left[ \kappa_0 (1-y) \right] \;.
\label{eq:13-56}
\end{equation}
Note that $\kappa_0 = -i \tilde{k}_0$ is the imaginary part
of $\tilde k_0$ and
$\kappa_0^2 = - \tilde{k}_0^2$.
As regards other solutions of~(\ref{eq:transc}) called
excited states, i.~e., those for
$\tilde k_m$ with $m>0$, nothing special happens at
$\Omega=-2$,
so that these wave numbers are always real.
The situation for ground state $m = 0$ at different values
of
dimensionless drift parameter $\Omega$ is summerized in
Table~\ref{tab1} which presents the solutions $\kappa_0$
from
transcendental equation~(\ref{eq:13-55}) together with
$\lambda_0 = - \, \kappa_0^2 + \Omega^2/4$
and $\tilde{k}_0$ from transcendental
equation~(\ref{eq:transc})
together with eigenvalues $\lambda_0 = \tilde{k}_0^2 +
\Omega^2/4$.
Table~\ref{tab2} shows the behaviour of lowest wave numbers
$\tilde{k}_m$
with $m=0, 1, \ldots, 5$. The results are plotted in
Fig.~\ref{kandlambda}.

\begin{table}
\caption{The ground--state wave number $\kappa_0$ (for
$\Omega \le -2$)
and $\tilde k_0$ (for $\Omega \ge -2$) and eigenvalue
$\lambda_0$ depending on the dimensionless drift parameter
$\Omega$.}
\begin{center}
\begin{minipage}[t]{\textwidth}
\hfill
\begin{tabular}{|@{\hspace{2ex}}c@{\hspace{2ex}}|
@{\hspace{2ex}}c@{\hspace{2ex}}|@{\hspace{2ex}}c@{\hspace{
2ex}}|}
\hline
\rule[-3mm]{0mm}{8mm}
$\Omega$  &  $\kappa_0$ & $\lambda_0$\\ \hline
\hline
$-9.00$   &  $4.499$  &   $0.010$ \\ \hline
$-8.50$   &  $4.248$  &   $0.015$ \\ \hline
$-8.00$   &  $3.997$  &   $0.021$ \\ \hline
$-7.50$   &  $3.745$  &   $0.031$ \\ \hline
$-7.00$   &  $3.493$  &   $0.045$ \\ \hline
$-6.50$   &  $3.240$  &   $0.064$ \\ \hline
$-6.00$   &  $2.984$  &   $0.091$ \\ \hline
$-5.50$   &  $2.726$  &   $0.128$ \\ \hline
$-5.00$   &  $2.464$  &   $0.178$ \\ \hline
$-4.50$   &  $2.195$  &   $0.245$ \\ \hline
$-4.00$   &  $1.915$  &   $0.333$ \\ \hline
$-3.50$   &  $1.617$  &   $0.446$ \\ \hline
$-3.00$   &  $1.288$  &   $0.591$ \\ \hline
$-2.50$   &  $0.888$  &   $0.774$ \\ \hline
$-2.00$   &  $0.000$  &   $1.000$ \\ \hline
\end{tabular}
\hfill
\begin{tabular}{|@{\hspace{2ex}}r@{\hspace{2ex}}|
@{\hspace{2ex}}c@{\hspace{2ex}}|@{\hspace{2ex}}c@{\hspace{
2ex}}|}
\hline
\rule[-3mm]{0mm}{8mm}
$\Omega$ & ${\tilde k}_0$ & $\lambda_0$\\ \hline
\hline
$-2.00$   &  $0.000$  &   $1.000$ \\ \hline
$-1.50$   &  $0.845$  &   $1.276$ \\ \hline
$-1.00$   &  $1.165$  &   $1.608$ \\ \hline
$-0.50$   &  $1.393$  &   $2.004$ \\ \hline
$ 0.00$   &  $1.571$  &   $2.468$ \\ \hline
$ 0.50$   &  $1.715$  &   $3.005$ \\ \hline
$ 1.00$   &  $1.836$  &   $3.623$ \\ \hline
$ 1.50$   &  $1.939$  &   $4.325$ \\ \hline
$ 2.00$   &  $2.028$  &   $5.116$ \\ \hline
$ 2.50$   &  $2.106$  &   $5.999$ \\ \hline
$3.00$    &  $2.174$  &   $6.979$ \\ \hline
$ 3.50$   &  $2.235$  &   $8.058$ \\ \hline
$ 4.00$   &  $2.288$  &   $9.239$ \\ \hline
$4.50$    &  $2.337$  &   $10.525$ \\ \hline
$5.00$    &  $2.381$  &   $11.917$ \\ \hline
\end{tabular}
\hfill
\mbox{}
\end{minipage}
\end{center}
\label{tab1}
\end{table}

\begin{figure}
\begin{center}
\includegraphics[scale=0.29]{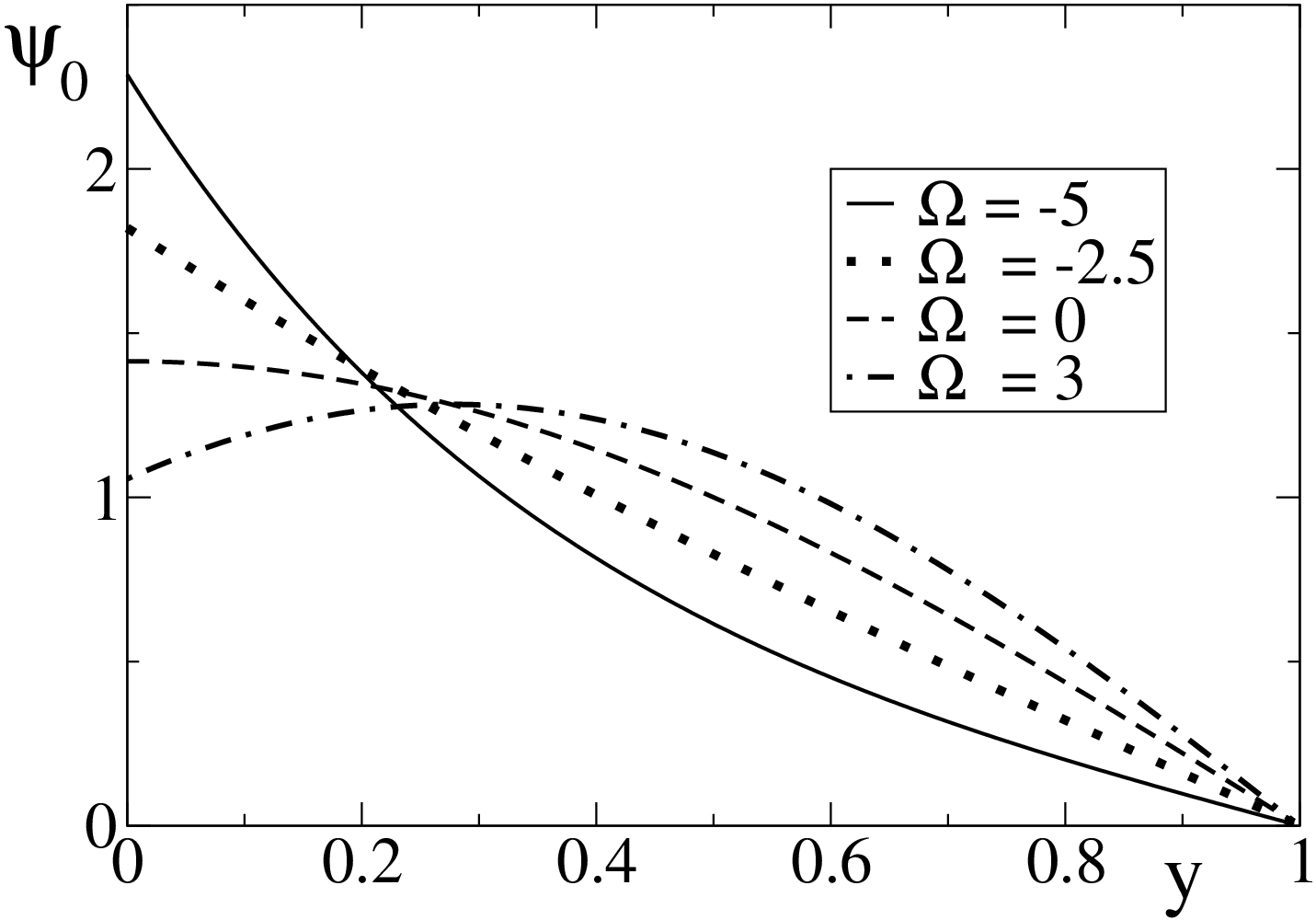}
\end{center}
\caption
{\label{h0}
The eigenfunction $\psi_0(y)$ for different values of
control parameter $\Omega$.}
\end{figure}

\begin{figure}
\begin{center}
\hspace*{\fill}
\includegraphics[scale=0.25]{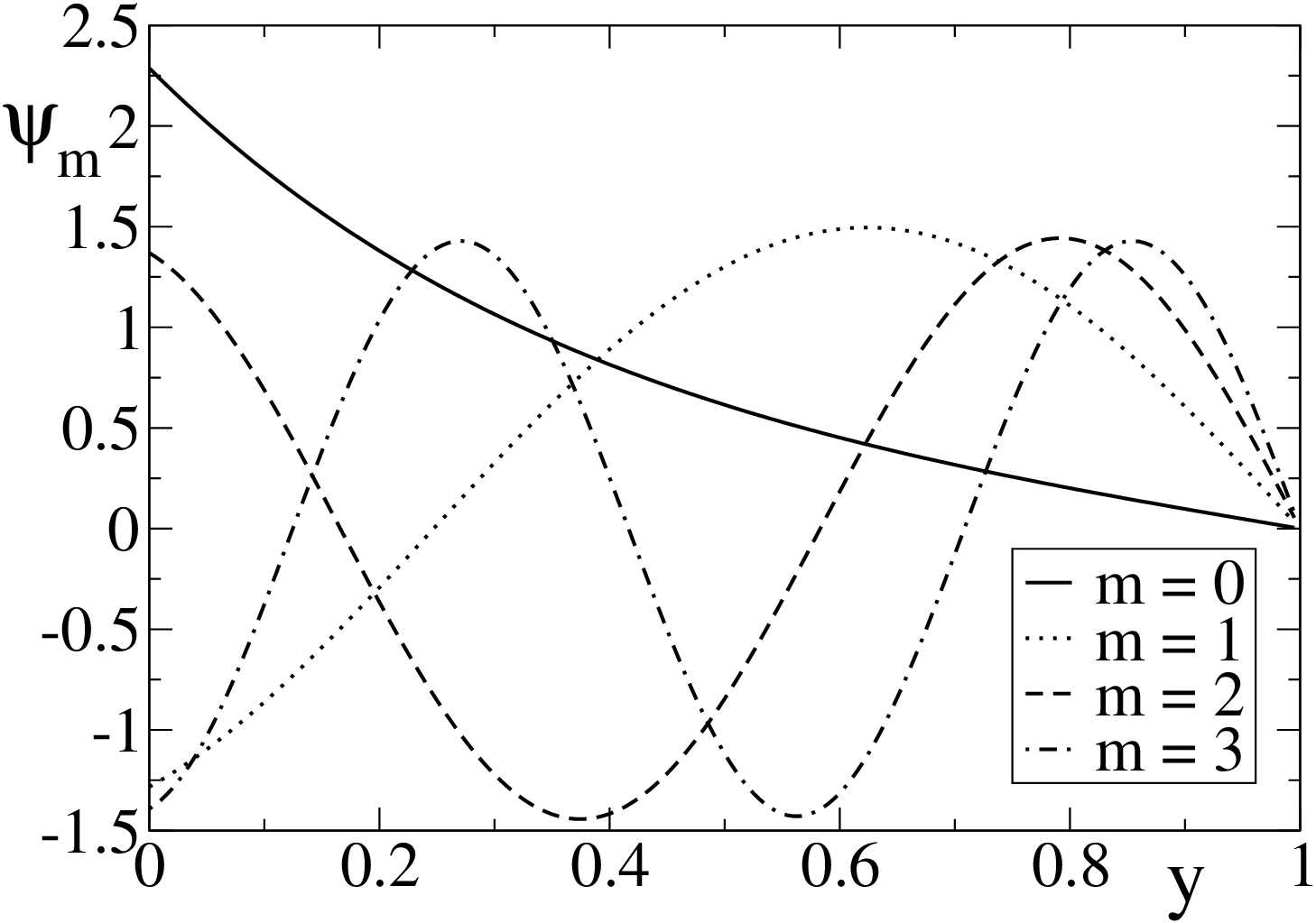}
\hspace*{\fill}
\includegraphics[scale=0.25]{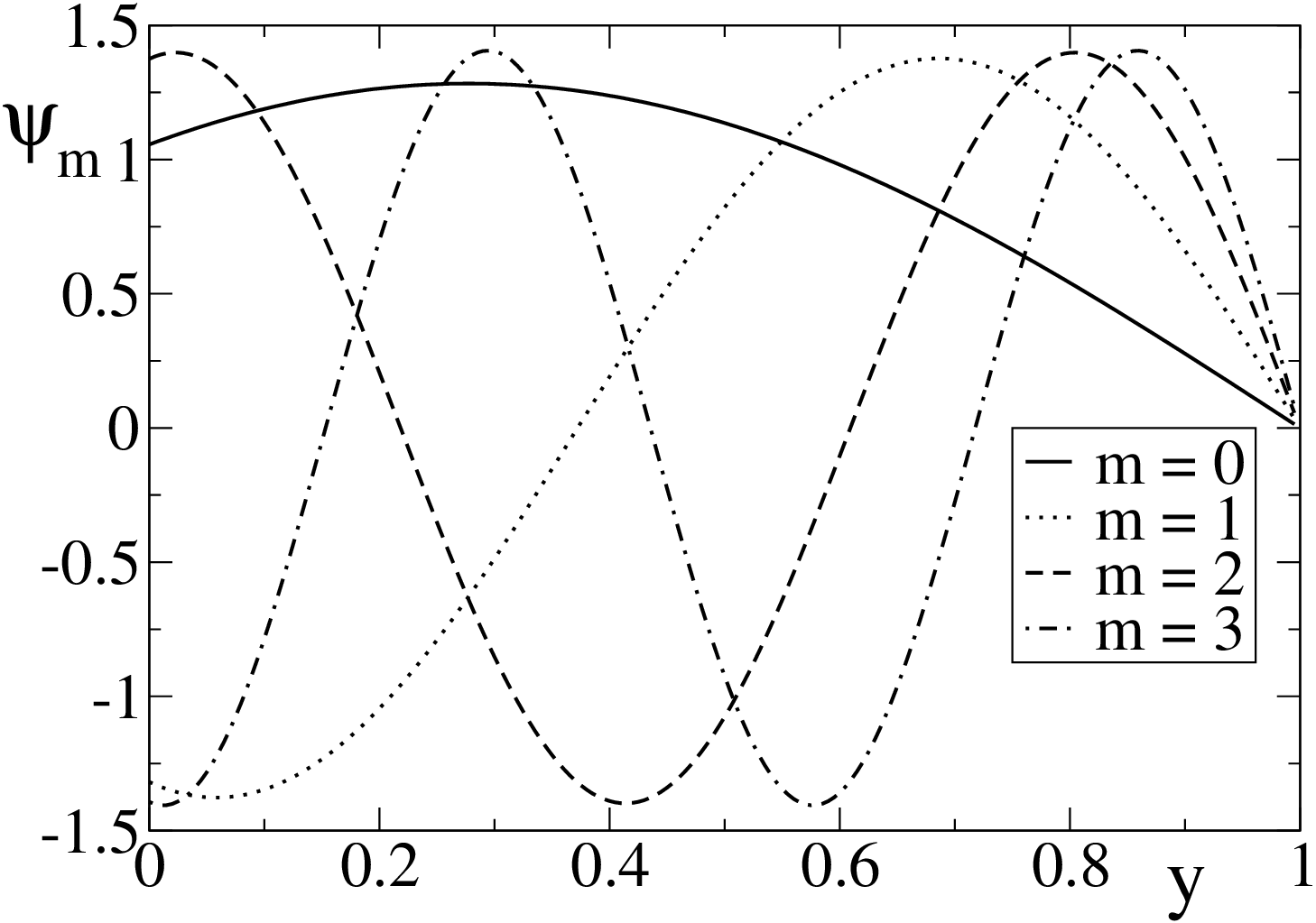}
\hspace*{\fill}
\end{center}
\caption
{\label{hm}
The eigenfunctions $\psi_m(y)$ for $m = 0, 1, 2, 3$ and
for $\Omega = - 5.0$ (left) and $\Omega = 3.0$ (right).}
\end{figure}

\begin{table}
\caption{The wave numbers $\tilde k_m$ ($m=0, 1, \ldots, 5$)
depending on the dimensionless drift parameter $\Omega$.}
\begin{center}
\begin{tabular}{
|c||c|c|c|c|c|c|c|c|c|}
\hline
\rule[-3mm]{0mm}{8mm}
$\Omega$  &  $-10.0$ & $-5.0$ & $-2.0$ &   $-1.0$ & $0.0$ & 
$1.0$ &      $2.0$ & $5.0$ & $10.0$\\ \hline \hline
$m = 0$ &  $4.999$  &   $2.464$   &  $0.000$   & $1.165$   &
$1.571$   & $1.836$ & $2.028$  & $2.381$ & $2.653$\\ \hline
$m = 1$ &  $3.790$  &   $4.172$   &  $4.493$   & $4.604$   &
$4.712$   & $4.816$   & $4.913$  & $5.163$ & $5.454$\\
\hline
$m = 2$ &  $7.250$  &   $7.533$  &  $7.725$   & $7.789$   &
$7.854$   & $7.917$   & $7.979$  & $8.151$ & $8.391$\\
\hline
$m = 3$ &  $10.553$ &   $10.767$  &  $10.904$  & $10.949$  &
$10.995$  & $11.040$  & $11.085$ & $11.214$ & $11.408$\\
\hline
$m = 4$ &  $13.789$ &   $13.959$  &  $14.066$  & $14.101$  &
$14.137$  & $14.172$  & $14.207$ & $14.310$ & $14.469$\\
\hline
$m = 5$ &  $16.992$ &   $17.133$  &  $17.220$  & $17.249$  &
$17.279$  & $17.308$  & $17.336$ & $17.421$ & $17.556$\\
\hline
\end{tabular}
\end{center}
\label{tab2}
\end{table}
In general (for arbitrary $\Omega$), the eigenfunctions are
orthogonal  and normalised, i.~e.,
\begin{equation}
\int_0^{1} \psi_l(y) \psi_m(y) dy = \delta_{ml}\;.
\end{equation}
Figure~\ref{h0} shows the ground eigenstate $(m=0)$ for
different
parameter values $\Omega$, whereas Fig.~\ref{hm} gives a
collection of
eigenstate functions $(m=0, 1, \ldots, 5)$ for $\Omega =
-5.0$ and
$\Omega=3.0$.

In the following, explicit formulae (where $\psi_m(y)$ is
specified) are
written for the case $\Omega > -2$.

\begin{center}
\begin{figure}
%\begin{center}
\hspace*{\fill}
\includegraphics[scale=0.25]{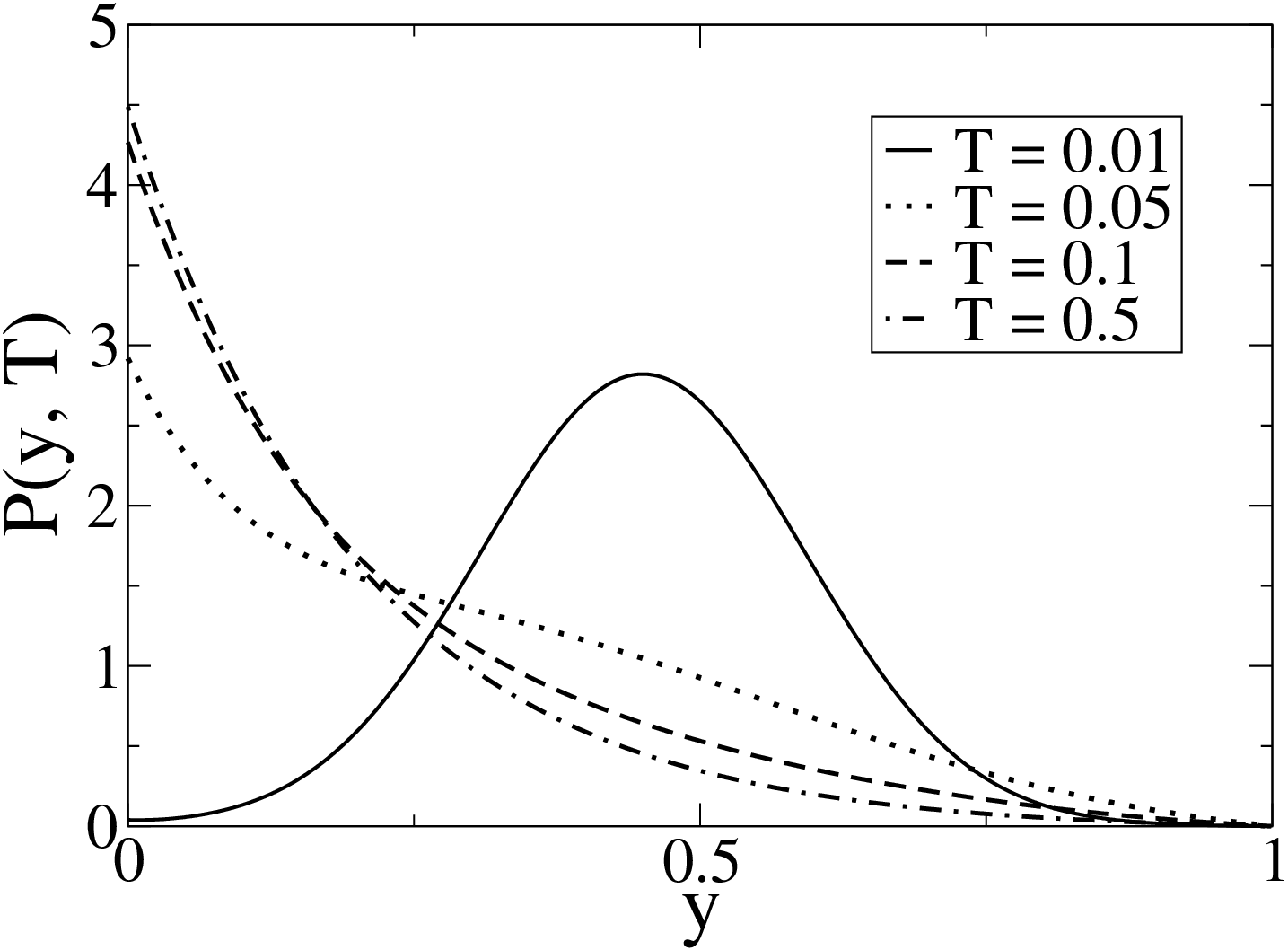}
\hspace*{\fill}
\includegraphics[scale=0.25]{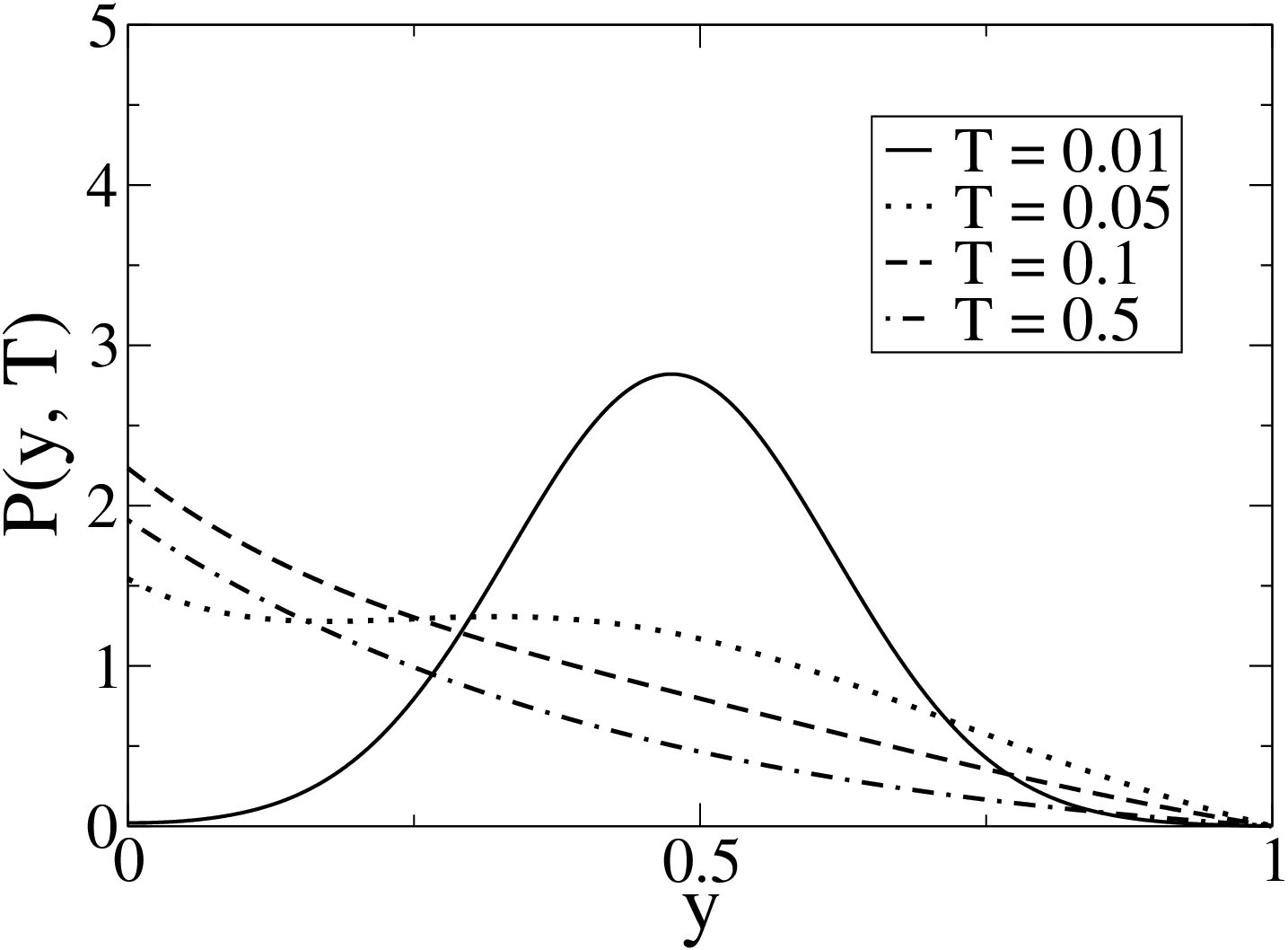}
\hspace*{\fill}
\\
\hspace*{\fill}
\includegraphics[scale=0.25]{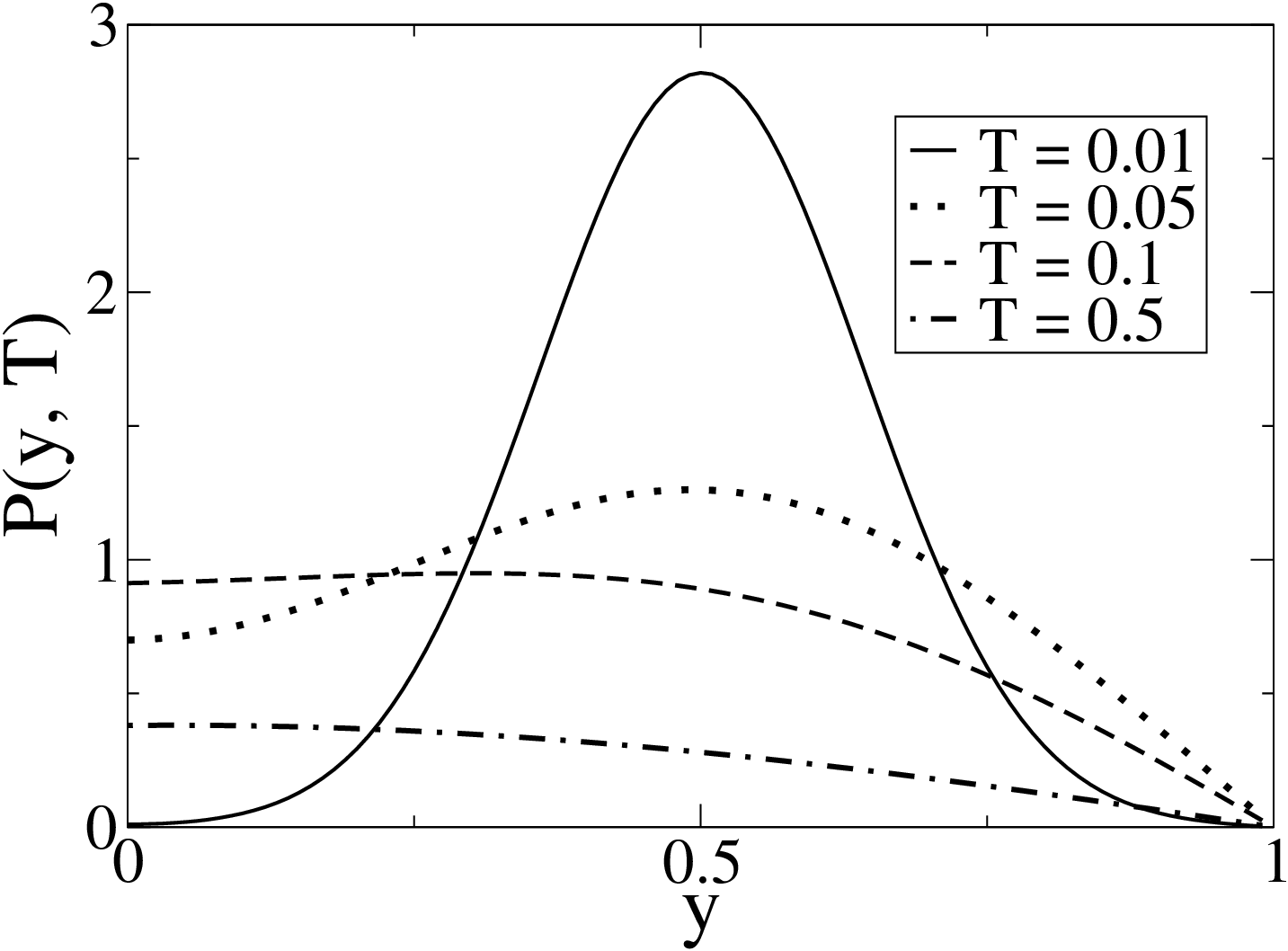}
\hspace*{\fill}
\includegraphics[scale=0.25]{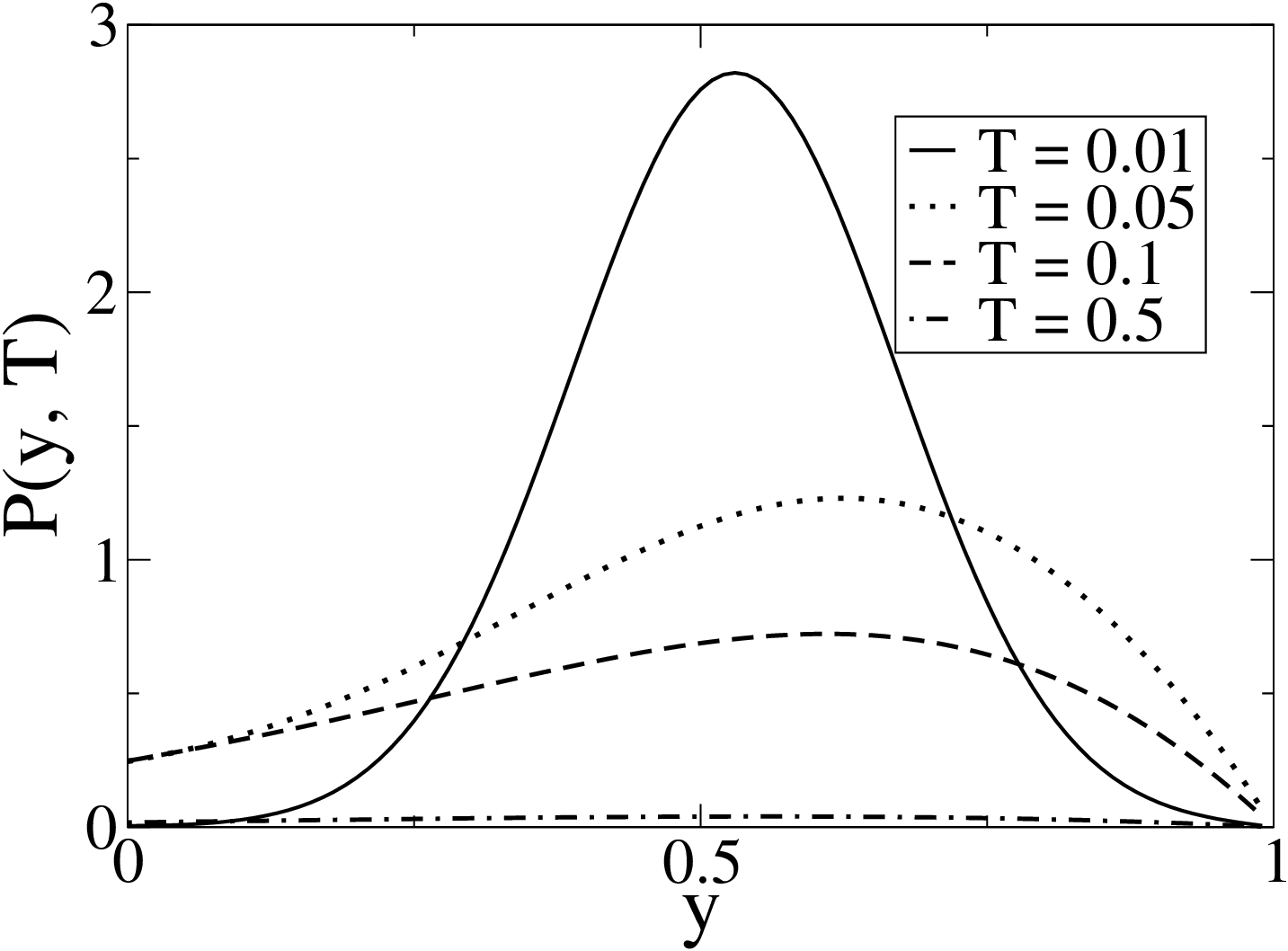}
\hspace*{\fill}
%\end{center}
\caption
{\label{prom2}
The solution of drift--diffusion Fokker--Planck equation
with initial condition $y_0=0.5$ for
different values of the control parameter $\Omega$, i.~e.
$\Omega = - 5.0$ (top left), $\Omega = -\,2.5$ (top right),
$\Omega = 0.1$ (bottom left), $\Omega = 3.0$ (bottom
right).}
\end{figure}
\end{center}

In order to construct the time--dependent solution for
$Q(y,t)$, which
fulfills the initial condition, we consider the
superposition of all
particular solutions with different eigenvalues $\lambda_m$
\begin{equation}
Q(y,T) =
\sum_{m=0}^\infty C_m e^{-\lambda_m T} \psi_m(y) \;.
\label{eq:super}
\end{equation}
By inserting the initial condition
\begin{equation}
P(y,T=0) = e^{\frac{\Omega}{2} y} Q(y,T=0) = \delta(y-y_0)
\end{equation}
into~(\ref{eq:super}) we obtain
\begin{equation}
\sum\limits_{m=0}^{\infty} C_m \psi_m(y) =
e^{-\frac{\Omega}{2} y}
\delta(y-y_0) \;.
\end{equation}
Now we expand the right hand side of this equation by using
the
basis of orthonormalised eigenfunctions~(\ref{eq:Hn}) and
identify $C_m$
with the corresponding coefficient at $\psi_m$, i.~e.,
\begin{equation}
C_m = \int e^{\frac{\Omega}{2} y} \delta(y-y_0) \psi_m dy
= e^{-\frac{\Omega}{2} y_0} \psi_m(y_0) \;.
\end{equation}
This allows us to write the solution for $P(y,T)$ as
\begin{equation}
P(y,T) = e^{\frac{\Omega}{2}(y-y_0)}
\sum\limits_{m=0}^{\infty}
e^{-\lambda_m T} \psi_m(y_0) \psi_m(y) \label{eq:pp}\;,
\end{equation}
with eigenfunctions (\ref{eq:Hn}) and (\ref{eq:13-56}) of
ground state
$(m=0)$
\begin{equation}
\psi_0(y) =
\left\lbrace
\begin{array}{cc}
\displaystyle{\sqrt{\frac{2}{1+ \frac{\Omega}{2}\,
\frac{1}{\tilde{k}_0^2+\Omega^2/4}} }}\,
\sin \left[ \tilde k_0 (1-y) \right]\,, &  \Omega >-\,2  \\
 & \\
\sqrt{3}\,\left(1 - y\right)\,,  &  \Omega = -\,2\\
 & \\
\displaystyle{\sqrt{-\,\frac{2}{1+ \frac{\Omega}{2} \,
\frac{1}{-\,\kappa_0^2+\Omega^2/4}} }}\,
\sinh \left[ \kappa_0 (1-y) \right]\,,  & \Omega < -\,2
\end{array} \right.
\end{equation}
and all other eigenfunctions (\ref{eq:Hn})
\begin{equation}
\psi_m(y)= \sqrt{\frac{2}{1+ \frac{\Omega}{2}\,
\frac{1}{\tilde{k}_m^2+\Omega^2/4}} }\,
\sin \left[ \tilde k_m (1-y) \right] \quad m = 1,2,\ldots \;
.
\end{equation}
The eigenvalue of ground state $(m=0)$ is given by
\begin{equation}
\lambda_0 = \left\lbrace
\begin{array}{cl}
{\tilde k}_0^2 + \Omega^2/4\,, & \Omega > -\,2 \\
 & \\
1\,, & \Omega = -\,2\\
 & \\
 -\,\kappa_0^2 + \Omega^2/4\,, & \Omega < -\,2
\end{array}\right.
\end{equation}
and all others are
\begin{equation}
\lambda_m = {\tilde k}_m^2 + \Omega^2/4 \qquad m =
1,2,\ldots \; ,
\end{equation}
where the wave numbers are calculated from transcendental
equation
(\ref{eq:13-48})
\begin{eqnarray}
\tilde{k}_0 & : & \quad \tan{\tilde k}_0 =
-\,\frac{2}{\Omega}\,{\tilde k}_0
\qquad \; \; \Omega > -\,2 \\
\kappa_0 & : & \quad \tanh \kappa_0 = -\,\frac{2}{\Omega}\,
\kappa_0
\qquad \; \Omega < -\,2 \\
\tilde{k}_m & : & \quad \tan{\tilde k}_m  =
-\,\frac{2}{\Omega}\,{\tilde k}_m
\qquad m = 1,2,\ldots \; .
\label{tr}
\end{eqnarray}
The set of Figures~\ref{prom2} illustrates the time
evolution of probability density (\ref{eq:pp}) choosing
different
parameter values
$\Omega$.
%

%%%%%%%%%%%%%%%%%%%%%%%%%%%%%%%%%%%%%%%%%%%%%%%%%%%%%%%%%%%%
%%

\section{First passage time probability density}

\begin{figure}
\begin{center}
\hspace*{\fill}
\includegraphics[scale=0.25]{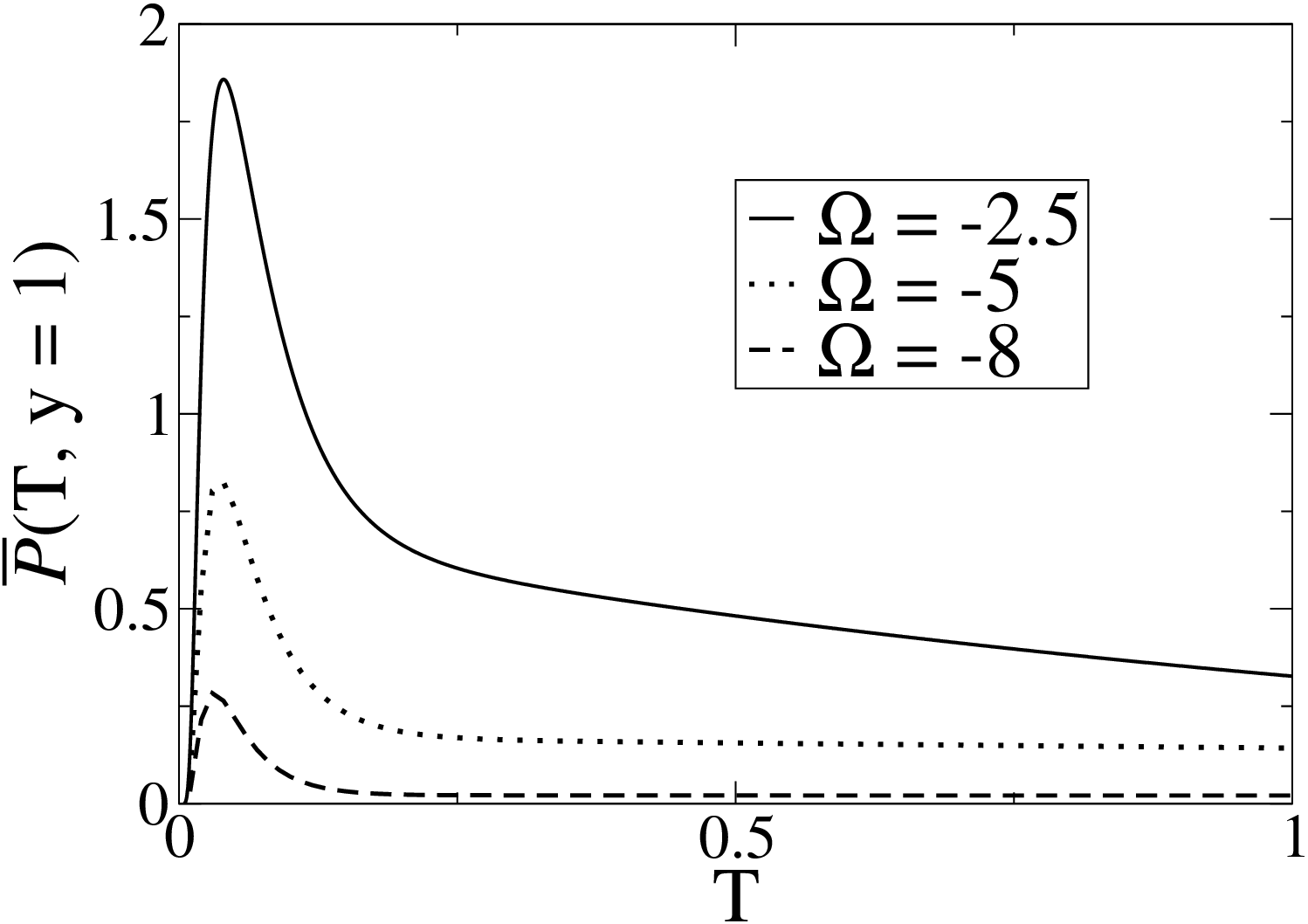}
\hspace*{\fill}
\includegraphics[scale=0.25]{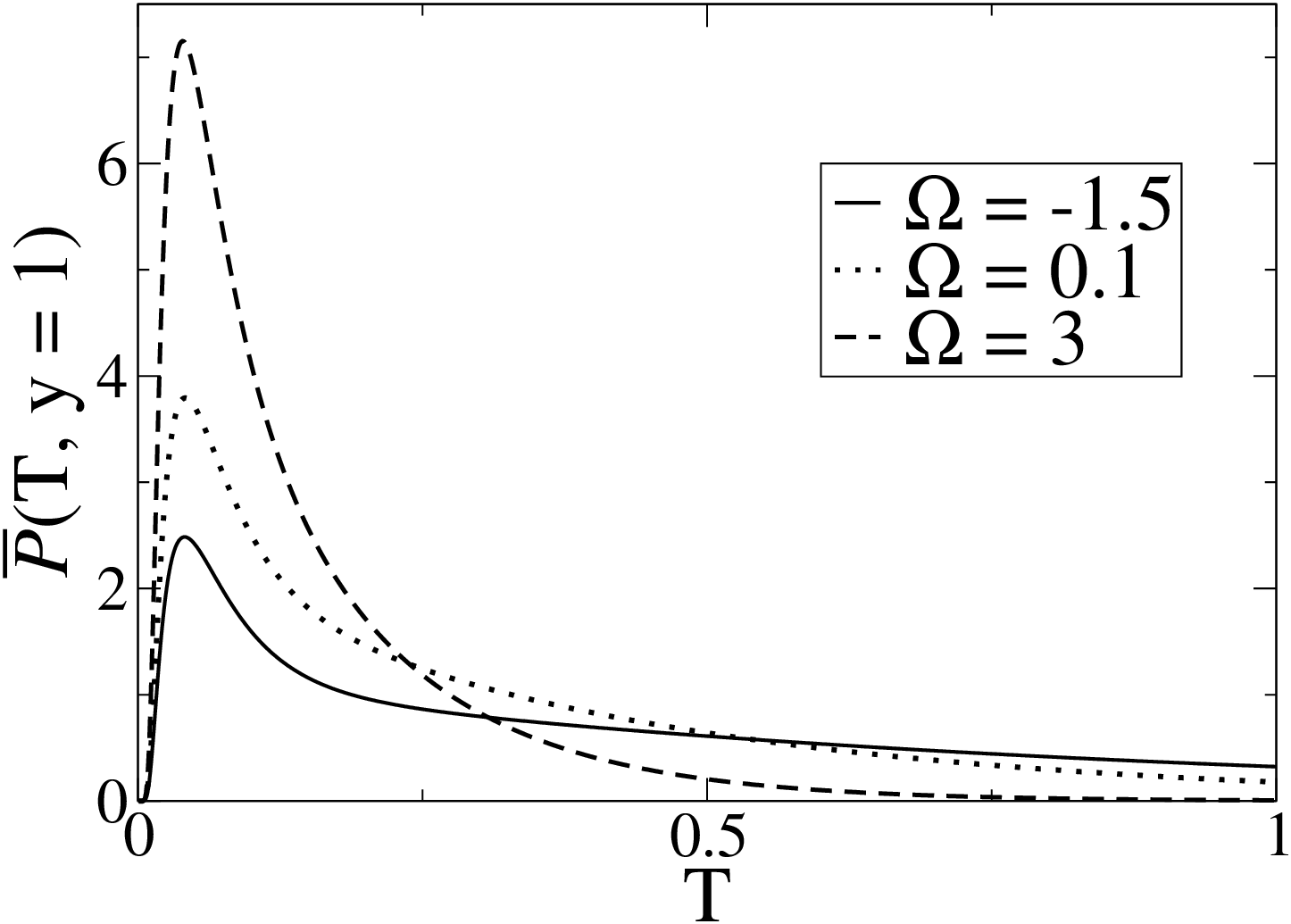}
\hspace*{\fill}
\end{center}
\caption
{\label{fptprom1}
The first passage time probability density distribution
$\overline{\mathcal{P}}(T,y=1)$ for $\Omega < - 2$ (left)
and
$\Omega > - 2$ (right).}
\end{figure}

\begin{figure}
\begin{center}
\includegraphics[scale=0.30]{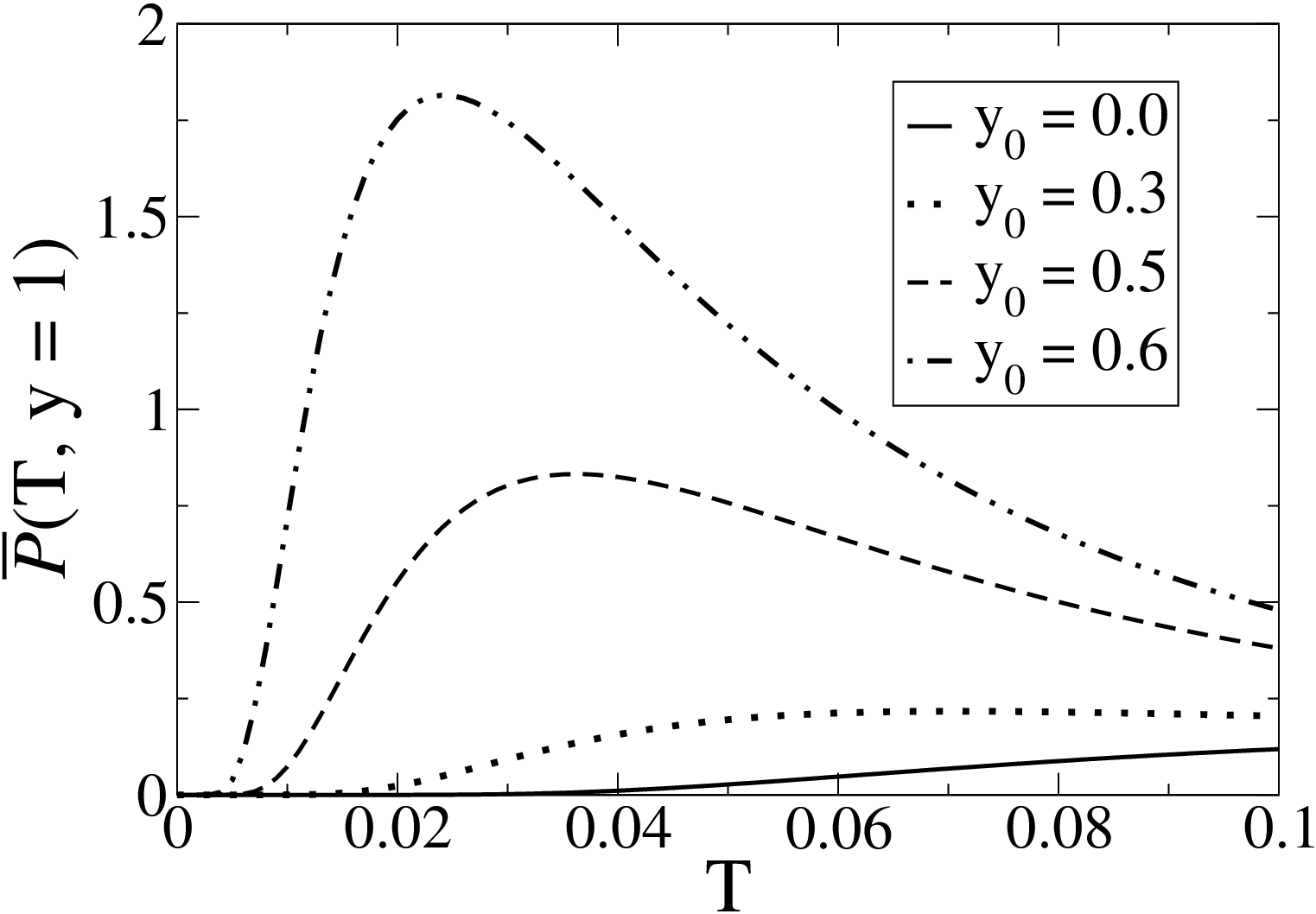}
\end{center}
\caption
{\label{fptprom1_y0}
Short time behaviour of first passage time probability
density distribution
$\overline{\mathcal{P}}(T,y=1)$ for different initial
conditions
$0 \le y_0 \le 1$ showing time lag.}
\end{figure}

It has been shown in previous sections that the
probability density $P(y,T)$ is not normalized
under given restrictions, i.~e. reflected at $y
=0$ and absorbed at $y=1$. Due to that fact, let us
apply here the balance equation in the open system given in
dimensionless variables by
\begin{equation}
\overline{\mathcal{P}}(T, y=1) =
- \frac{\partial}{\partial T} \int\limits_0^1 P(y,T) \, dy
\label{eq:13.15}
\end{equation}
which relates the probability $P(y,T)$ that the system is
still in a
state $y \in
[0,1]$ with the probability flux $\overline{\mathcal
P}(T,y=1)$ out
of this
interval at the right absorbing boundary $y=1$ at time
moment $T$.
Hence, $\overline{\mathcal{P}}(T,y=1)$ is the first passage
time probability density~\cite{phr2,phr1,red}. It can
be calculated by using obtained results of previous section.
The first passage time probability density distribution 
$\overline{\mathcal{P}}$ (breakdown probability density)
depending on $\Omega$ reads as follows
\begin{enumerate}
\item $\Omega > -\,2 \quad$
\begin{equation}\label{haha1}
\overline{\mathcal{P}}(T,y=1) = 2
e^{\frac{\Omega}{2}(1-y_0)} \sum\limits_{m=0}^{\infty}
\frac{ e^{-\left( \tilde k_m^2 + \Omega^2/4 \right) T}}
{1 + \frac{\Omega}{2} \frac{1}{\tilde k_m^2 + \Omega^2/4} }
\tilde k_m \sin \left[ \tilde k_m(1-y_0) \right]
\end{equation}
\item $\Omega = -\,2$
\begin{eqnarray}\label{haha2}
\overline{\mathcal{P}}(T,y=1) &=&  e^{-\,(1-y_0)}
\Bigg[ 3\,(1-y_0)\,e^{-\,T}  \nonumber\\
&+& \left. 2\,\sum\limits_{m=1}^{\infty}
\frac{ e^{-\left( \tilde k_m^2 + 1 \right) T}}
{1 - \frac{1}{\tilde k_m^2 + 1} }
\tilde k_m \sin \left[ \tilde k_m(1-y_0) \right]\right]
\end{eqnarray}
\item $\Omega < -\,2$
\begin{eqnarray}
\overline{\mathcal{P}}(T,y=1) &  = &
2 e^{\frac{\Omega}{2}(1-y_0)} \nonumber \label{eq:13.103}\\
&\times & \left[-\, \frac{e^{-\left( -\,\kappa_0^2 +
\Omega^2/4 \right) T}}
{1 + \frac{\Omega}{2} \frac{1}{-\,\kappa_0^2 + \Omega^2/4} }
\kappa_0 \sinh \left[ \kappa_0(1-y_0) \right]\right.
\nonumber \\
& + & \left.\sum\limits_{m=1}^{\infty}
\frac{ e^{-\left( \tilde k_m^2 + \Omega^2/4 \right) T}}
{1 + \frac{\Omega}{2} \frac{1}{\tilde k_m^2 + \Omega^2/4} }
\tilde k_m \sin \left[ \tilde k_m(1-y_0) \right]\right]
\label{eq:13.105}
\end{eqnarray}
\end{enumerate}
The outflow distribution $\overline{\mathcal{P}}(T,y=1)$ is
shown in
Fig.~\ref{fptprom1} (with different values of dimensionless
drift $\Omega$)
as well as in Fig.~\ref{fptprom1_y0} (with different values
of initial
condition $y_0$).
%

%%%%%%%%%%%%%%%%%%%%%%%%%%%%%%%%%%%%%%%%%%%%%%%%%%%%%
\section{Cumulative breakdown probability}

\begin{figure}
\begin{center}
\hspace*{\fill}
\includegraphics[scale=0.25]{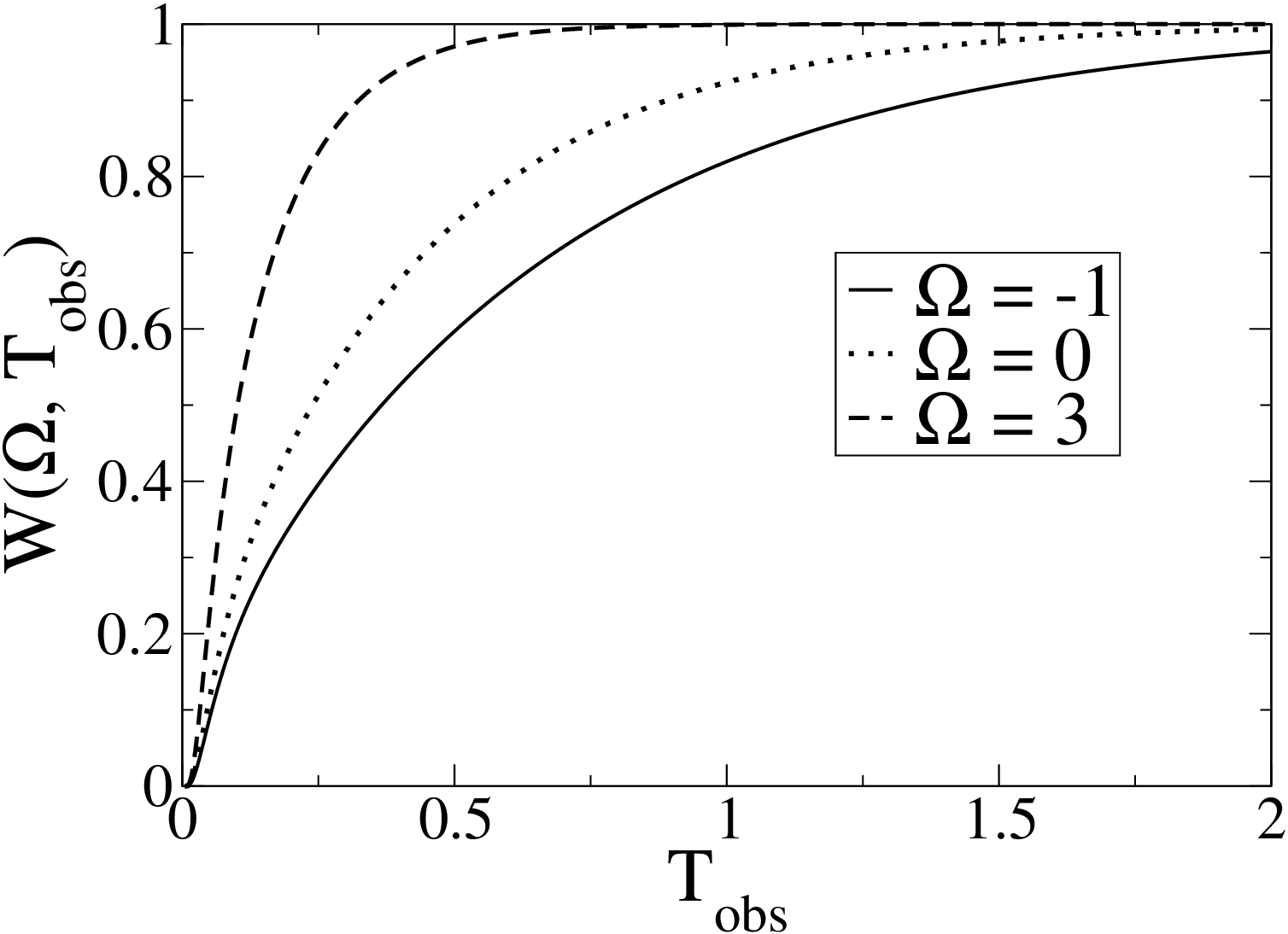}
\hspace*{\fill}
\includegraphics[scale=0.25]{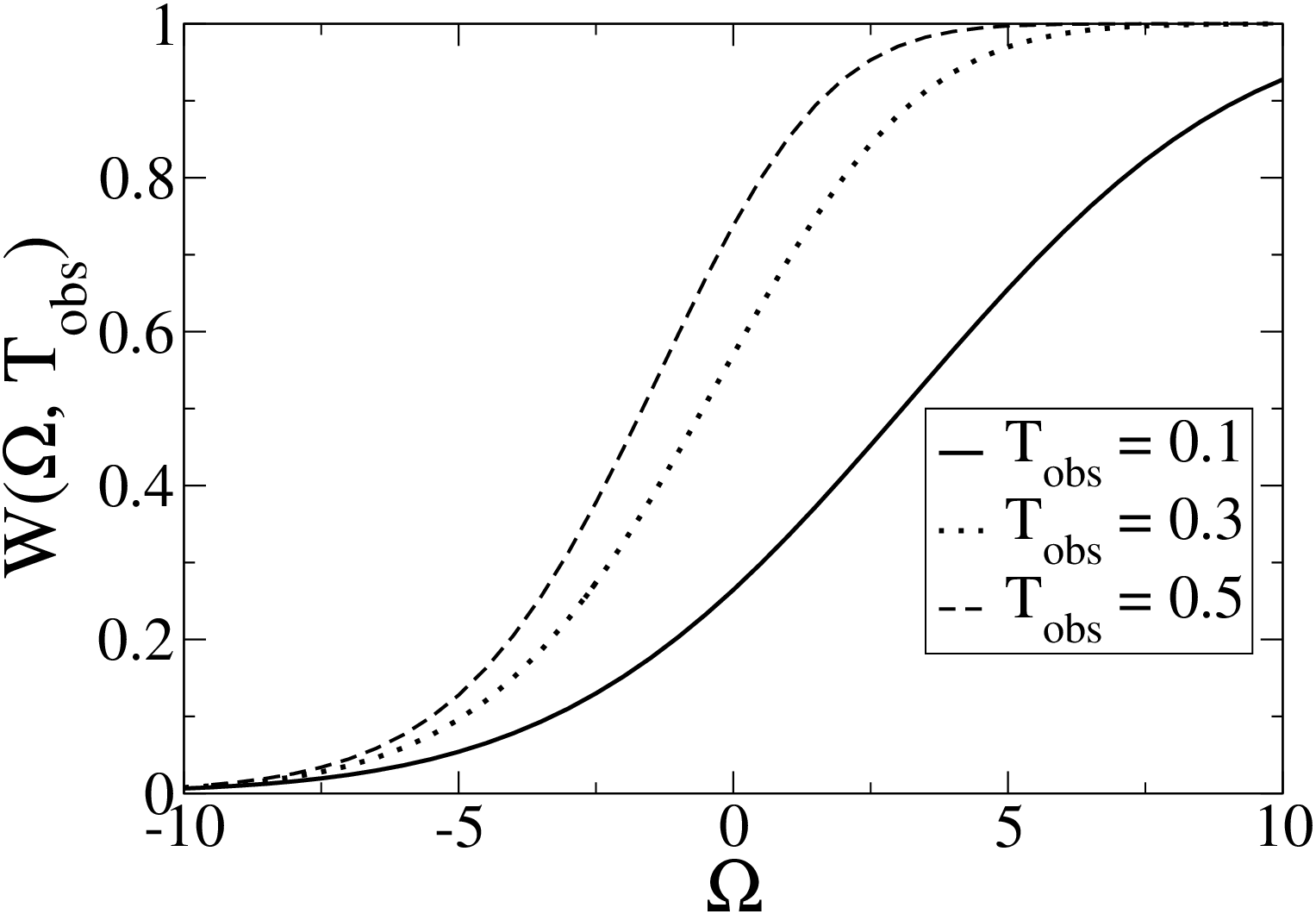}
\hspace*{\fill}
\end{center}
\caption{
The probability $W(\Omega, T_{obs})$ (\ref{eq:W})
 as function of observation time $T_{obs}$ with fixed
$\Omega$ (left) and vice versa (right).} \label{cum}
\end{figure}

The probability that the absorbing boundary $y=1$ is reached
within
certain observation time interval $0 \le T \le T_{obs}$ is
given by the
cumulative (breakdown) probability
\begin{equation}
W \left( \Omega,T=T_{obs} \right) = \int\limits_0^{T_{obs}}
\overline{\mathcal{P}}(T,y=1) \, dT
\label{eq:W}
\end{equation}
with $\overline{\mathcal{P}}(T,y=1)$ from
(\ref{eq:13.15}).
For $T_{obs} \to \infty$ we have $W \to 1$. Generally, we
obtain
\begin{enumerate}
\item $\Omega > -\,2 \quad$
\begin{equation}
W \left( \Omega,T_{obs} \right)= 2
e^{\frac{\Omega}{2}(1-y_0)} \sum\limits_{m=0}^{\infty}
\frac{ 1 - e^{-\left( \tilde k_m^2 + \Omega^2/4 \right)
T_{obs}}}
{ \tilde k_m^2 + \Omega^2/4 + \Omega/2 }
\tilde k_m \sin \left[ \tilde k_m(1-y_0) \right]
\end{equation}
\item $\Omega = -\,2$
\begin{eqnarray}
W \left( \Omega,T_{obs} \right) & = & e^{-\,(1-y_0)}
%\nonumber \\
%&\times &
\Bigg[ 3\,\left(1 - e^{-T_{obs}}\right)(1 - y_0) \nonumber
\\
& + & 2\,\sum\limits_{m=1}^{\infty}
\frac{1 - e^{-\left( \tilde k_m^2 + 1\right) T_{obs}}}
{\tilde k_m}
\sin \left[ \tilde k_m(1-y_0) \right]\Bigg]\,.
\end{eqnarray}
\item $\Omega < -\,2$
\begin{eqnarray}
W \left( \Omega,T_{obs} \right) & = &  2 \,
e^{\frac{\Omega}{2}(1-y_0)}\nonumber \\
&\times & \left[-\, \frac{1  - e^{-\left(-\kappa_0^2 +
\Omega^2/4 \right) T_{obs}}}
{ -\,\kappa_0^2 + \Omega^2/4 + \Omega/2 }
\kappa_0 \sinh \left[ \kappa_0(1-y_0) \right]\right. \\
& + & \left.\sum\limits_{m=1}^{\infty}
\frac{ 1  - e^{-\left( \tilde k_m^2 + \Omega^2/4 \right)
T_{obs}}}
{ \tilde k_m^2 + \Omega^2/4 + \Omega/2 }
\tilde k_m \sin \left[ \tilde k_m(1-y_0) \right]\right]
\nonumber
\end{eqnarray}
\end{enumerate}
Figure~\ref{cum} shows $W \left( \Omega,T_{obs} \right)$ as
a function of observation
time $T_{obs}$
(left) as well as parameter dependence $\Omega$ (right).
%
%%%%%%%%%%%%%%%%%%%%%%%%%%%%%%%%%%%%%%%%%%%%%%%%%%%%

\section{Limit case for large positive values of the control
parameter}

Consider parameter limit $\Omega \to +\infty$ which
corresponds either
to large positive drift $v$ and/or large interval $b-a$, or
to a small
diffusion coefficient $D$. In this
case, for a given $m$, the solution of the transcendental
equation can be
found in the form $\tilde k_m = \pi (m+1) - \varepsilon_m$,
where
$\varepsilon_m$ is small and positive. From the periodicity
property we
obtain
\[
\cos \tilde k_m = \cos(\pi (m+1) -\varepsilon_m)=
-(-1)^m \cos(\varepsilon_m) = -(-1)^m + {\mathcal O} \left(
\varepsilon_m^2 \right)
\]
\[
\sin \tilde k_m = \sin(\pi (m+1) -\varepsilon_m)=
(-1)^m \sin(\varepsilon_m)= (-1)^m \varepsilon_m
+ {\mathcal O} \left( \varepsilon_m^3 \right) \;.
\]
By inserting this into the transcendental
equation~(\ref{eq:transc}), we obtain
\begin{eqnarray}
\varepsilon_m &=& \frac{2}{\Omega} \pi (m+1) + {\mathcal O}
\left( \Omega^{-2}
\right) \, , \\
\sin(\tilde k_m) &=& \frac{2}{\Omega} (-1)^m \pi (m+1)
+ {\mathcal O} \left( \Omega^{-2} \right) \; .
\end{eqnarray}
In this approximation the normalisation integral for large
$\Omega$
and the initial condition $y_0 \to 0$ can be written as
\begin{eqnarray}
I &=& \int\limits_0^{\infty} \overline{\mathcal{P}}(T,y=1)
\, dT =
2 e^{\Omega/2} \sum\limits_{m=0}^{\infty} \frac{\tilde
k_m \sin \left(\tilde k_m \right)}{\lambda_m + \Omega/2} \\
&\simeq& e^{\Omega/2} \sum\limits_{m=1}^{\infty}
\frac{-4}{\Omega} \frac{(-1)^m (\pi m)^2}{\pi^2 m^2 +
\Omega^2/4}
= e^{\Omega/2} \sum\limits_{m=-\infty}^{\infty}
\frac{-2}{\Omega} \frac{(-1)^m (\pi m)^2}{\pi^2 m^2 +
\Omega^2/4} \; .
\nonumber
\end{eqnarray}
Further on we set $(-1)^m= e^{i \pi m}$ and, in a continuum
approximation, replace the sum by the integral
\begin{equation}
I \simeq e^{\Omega/2} \int\limits_{-\infty}^{\infty}
\frac{-2}{\Omega} \frac{e^{i \pi m} (\pi m)^2}{\pi^2 m^2
+ \Omega^2/4} dm \;.
\label{eq:13.88}
\end{equation}
Now we make an integration contour in the complex plane,
closing it in
the upper plane ($\rm{Im} \, m >0$) at infinity
where $|e^{i \pi m}|$ is
exponentially small. According to the residue theorem, it
yields
\begin{equation}
I = 2 \pi i \sum_i \rm{Res}(m_i) = 2 \pi i
\rm{ Res}(m_0) \;,
\end{equation}
where $m_0= \frac{i \Omega}{2 \pi}$ is the location of the
pole in the
upper plane, found as a root of the equation
$\pi^2 m^2 + \Omega^2/4 =0$. According to the well--known
rule,
the residue is calculated by setting $m=m_0$ in the
enumerator
of~(\ref{eq:13.88}) and replacing the denominator with its
derivative at
$m=m_0$. It gives the desired result $I=1$, i.~e., the
considered
approximation gives correct normalisation of outflow
probability
density $\overline{\mathcal{P}}(T,y=1)$ at the right
boundary.

The probability distribution function $P(y,T)$ given by
(\ref{eq:pp})
can also be calculated in
such a continuum approximation. In this case the increment
of wave
numbers is
\begin{equation}
\Delta \tilde k_m = \tilde k_{m+1} - \tilde k_m = \pi +
\varepsilon_m - \varepsilon_{m+1}
\simeq \pi \left(1- \frac{2}{\Omega} \right)
\simeq \frac{\pi}{1+ 2 / \Omega} \;.
\end{equation}
Note that in this approximation for $\Omega \to \infty$ the
normalisation
constant $R_m$ in~(\ref{eq:Rm}) is related to the
increment
$\Delta \tilde k$ via
\begin{equation}
R_m^2  = \frac{2}{1+ \frac{\Omega}{2} \frac{1}{\tilde k_m^2
+
\Omega^2/4} } \simeq \frac{2}{1 + 2/\Omega} \simeq
\frac{2}{\pi} \Delta
\tilde k_m \;.
\end{equation}
Hence, the equation~(\ref{eq:pp}) for the probability
density can be written
as
\begin{eqnarray}
\label{eq:ppp}
&&\hspace{-4ex} P(y,T) = 2 e^{\frac{\Omega}{2} (y-y_0)}
\sum_{m=0}^\infty R_m^2 e^{-\lambda_m T}
\sin\left[\tilde{k}_m \left(1- y_0\right) \right]
\sin\left[\tilde{k}_m \left(1- y\right) \right] \\
&&\hspace{-4ex}\simeq
\frac{2}{\pi} e^{\frac{\Omega}{2} (y-y_0)}
\sum_{m=0}^\infty e^{-\left(\tilde{k}_m^2 + \Omega^2/4
\right)T}
\sin\left[\tilde{k}_m \left(1- y_0\right) \right]
\sin\left[\tilde{k}_m \left(1- y\right) \right] \Delta
\tilde k_m
\nonumber \;.
\end{eqnarray}
In the continuum approximation we replace the sum by the
integral
\begin{eqnarray}
&&\hspace{-5ex} P(y,T) \simeq
\frac{2}{\pi} e^{\frac{\Omega}{2} (y-y_0)}
\int\limits_0^{\infty} e^{-\left(\tilde{k}^2 + \Omega^2/4
\right)T}
\sin\left[\tilde{k} \left(1- y_0\right) \right]
\sin\left[\tilde{k} \left(1- y\right) \right]\! d \tilde k
\\
&&\hspace{-5ex}=
\frac{1}{\pi} e^{\frac{\Omega}{2} (y-y_0)}
\int\limits_0^{\infty} e^{-\left(\tilde{k}^2 + \Omega^2/4
\right)T}
\left( \cos \left[\tilde{k} \left(y- y_0\right) \right]
- \cos \left[\tilde{k} \left(2- y -y_0 \right) \right]
\right) d \tilde k
\nonumber \;.
\end{eqnarray}
In the latter transformation we have used the
identity\linebreak
$\sin \alpha \, \sin \beta = \frac{1}{2} \left( \cos( \alpha
-\beta) -
\cos( \alpha + \beta) \right)$. The resulting known
integrals yield
\begin{equation}
P(y,T) \simeq \frac{1}{\sqrt{4 \pi T}} e^{\frac{\Omega}{2}
\left(y-y_0- \frac{\Omega}{2}T \right)}
\left[ e^{-\frac{(y-y_0)^2}{4T}}
- e^{-\frac{(2-y-y_0)^2}{4T}} \right] \;.
\label{eq:pcont}
\end{equation}
\begin{figure}
\begin{center}
\includegraphics[scale=0.28]{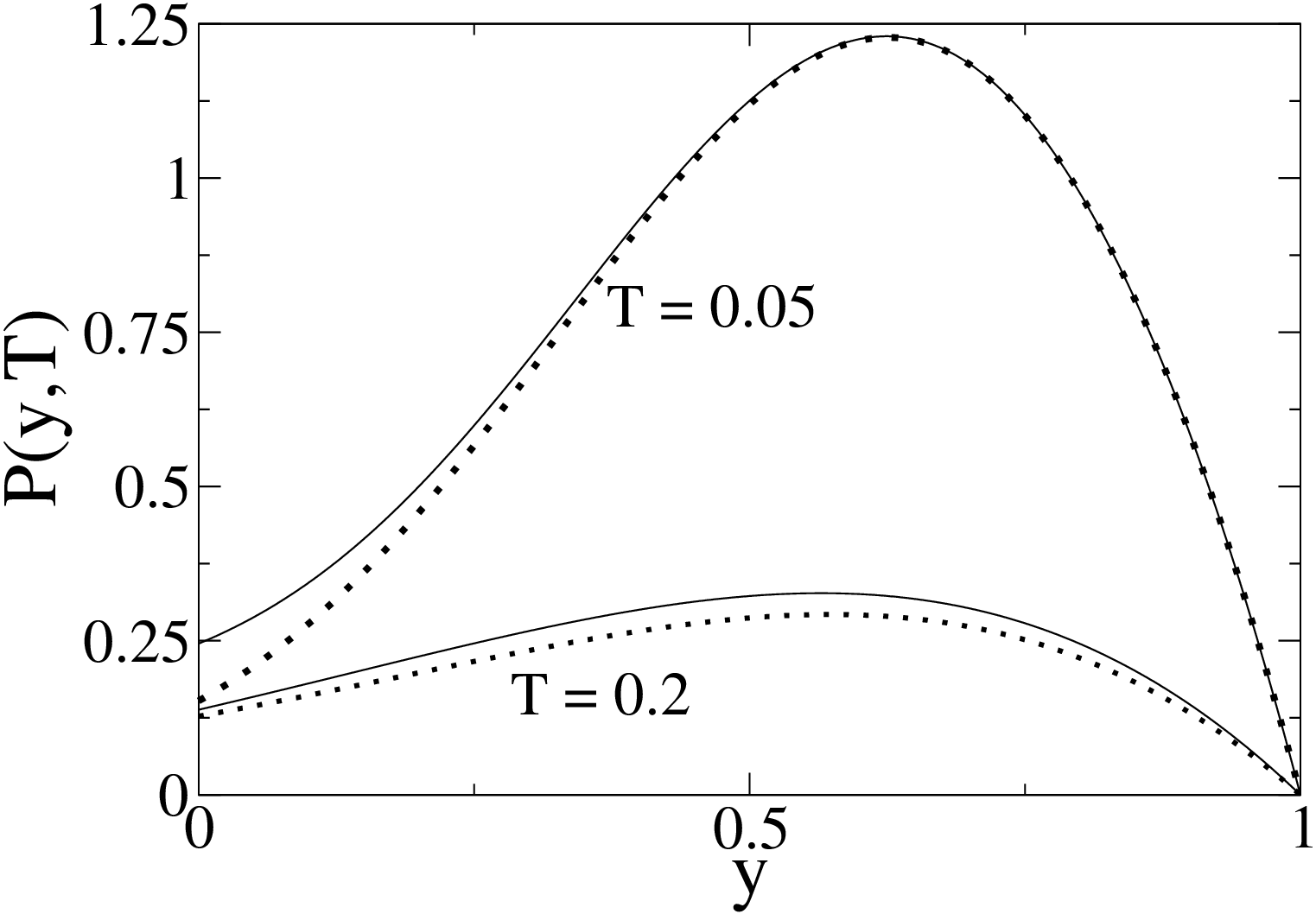}
\end{center}
\caption
{\label{compare}
Comparison of probability density $P(y,T)$ in
drift--diffusion--dynamics with finite boundaries
for two time moments. Parameter value is
$\Omega = 3.0$; initial condition is $y_0 = 0.5$.
The solid lines represent the exact result (\ref{eq:pp});
dotted lines
display the approximation (\ref{eq:pcont}).
}
\end{figure}
%The comparison between exact distribution (\ref{eq:pp}
The approximation (\ref{eq:pcont}) is shown in
Fig.~\ref{compare}.
For short enough times $4T \ll (2-y-y_0)^2$ the second term
is
very small. Neglecting this term, Eq.~(\ref{eq:pcont})
reduces to the known exact solution for natural boundary
conditions.

Based on~(\ref{eq:pcont}), it is easy to calculate the
probability flux
\begin{equation}
J(y,T)= \Omega P(y,T) - \frac{\partial}{\partial y} P(y,T)
\end{equation}
and the first passage time distribution
$\overline{\mathcal{P}}(T)=J(y=1,T)$
which takes a particularly simple form
\begin{equation}
\overline{\mathcal{P}}(T) = \frac{1-y_0}{\sqrt{4 \pi T^3}}
e^{-\frac{(1-y_0- \Omega T)^2}{4T}} \;.
\end{equation}
The cumulative breakdown probability~(\ref{eq:W}) is then
\begin{equation}
W(\Omega,T=T_{obs}) = \int\limits_0^{T_{obs}}
\frac{1-y_0}{\sqrt{4 \pi T^3}}
e^{-\frac{(1-y_0- \Omega T)^2}{4T}} dT \;.
\end{equation}
%%%
\section{Relationsship to Sturm--Liouville theory}
The particular drift--diffusion--problem over a finite
interval with
reflecting (left) and
absorbing (right) boundaries belongs to the following
general mathematical
theory named after Jacques Charles Francois Sturm
(1803--1855)
and Joseph Liouville (1809--1882).

The classical Sturm--Liouville theory considers a real
second--order
linear differential equation of the form~\cite{sturm}
\begin{equation} \label{sl01}
- \frac{d}{dx}\left[ p(x) \frac{d \psi}{dx} \right] + q(x)
\psi 
= \lambda \, w(x) \psi
\end{equation}
together with boundary conditions at the ends of interval
$\left[a,b\right]$ given by
\begin{eqnarray}
\alpha_1 \psi(x=a) + \alpha_2 \left. \frac{d \psi}{dx}
\right|_{x=a} 
&=&0 \; , \label{sl02} \\
\beta_1 \psi(x=b) + \beta_2 \left. \frac{d \psi}{dx}
\right|_{x=b} 
&=&0 \; . \label{sl03}
\end{eqnarray}

The particular functions $p(x), q(x), w(x)$ are real and
continuous on the
finite interval $\left[a,b\right]$ together with specified
values at the
boundaries. The aim of the Sturm--Liouville problem is to
find the values
of $\lambda$ (called eigenvalues $\lambda_n$) for which
there exist
non--trivial solutions of the differential
equation~(\ref{sl01}) satisfying
the boundary conditions~(\ref{sl02}) and (\ref{sl03}). The
corresponding
solutions (for such $\lambda_n$) are called eigenfunctions
$\psi_n(x)$ of
the problem.

Defining the Sturm--Liouville differential operator over the
unit interval
$\left[ 0, 1 \right]$ by
\begin{equation}
\mathcal{L} \psi =
- \frac{d}{dx}\left[ p(x) \frac{d \psi}{dx} \right] + q(x)
\psi 
\end{equation}
and putting the weight $w(x)$ to unity $(w=1)$ the general
equation~(\ref{sl01}) can precisely be written as eigenvalue
problem
\begin{equation}
\mathcal{L} \psi = \lambda \psi
\end{equation}
with boundary conditions ~(\ref{sl02})($a=0$) and
(\ref{sl03})
($b=1$) written as
\begin{equation}
\mathcal{B}_0 \psi = 0 \quad \; \quad \mathcal{B}_1 \psi = 0
\; .
\end{equation}

Assuming a differentiable positive function $p(x) > 0$ the
Sturm--Liouville operator is called regular and it is
self--adjoint to
fulfil
\begin{equation}
\int_0^1 \mathcal{L}\psi_1 \cdot \psi_2 =
\int_0^1 \psi_1 \cdot \mathcal{L}\psi_2 \; .
\end{equation}
Any self--adjoint operator has real nonnegative eigenvalues 
$\lambda_0 < \lambda_1 < \ldots < \lambda_n < \ldots \to
\infty$.
The corresponding eigenfunctions $\psi_n(x)$ have exact $n$
zeros in
$(0,1)$ and form an orthogonal set
\begin{equation} \label{sl08}
\int_0^1 \psi_n(x) \psi_m(x) dx = \delta_{mn} \; .
\end{equation}

The eigenvalues $\lambda_n$ of the classical
Sturm--Liouville problem
(\ref{sl01}) with positive function $p(x)>0$ as well as
positive weight
function $w(x)>0$ together with separated boundary
conditions (\ref{sl02})
and (\ref{sl03}) can be calculated by the following
expression
\begin{eqnarray}
\lambda_n \int_a^b \psi_n(x)^2 w(x) dx 
&=& \int_a^b \left[ p(x) \left(d\psi_n(x)/dx\right)^2 
+ q(x) \psi_n(x)^2\right] dx \nonumber \\
&& - \; \Big| p(x) \psi_n(x)
\left(d\psi_n(x)/dx\right)\Big|_a^b \; .
\label{sl09}
\end{eqnarray}

The eigenfunctions are mutually orthogonal $(m \neq n)$ and
usually normalized $(m=n)$
\begin{equation} \label{sl10}
\int_a^b \psi_n(x) \psi_m(x) w(x) dx = \delta_{mn}
\end{equation}
known as orthogonality relation (similar to (\ref{sl08}).

Comming back to the original drift--diffusion problem
written
in dimensionless variables over unit interval $0 \le y \le
1$ and
recalling (\ref{eq:13.34} the separation constant $\lambda$
appears in
the following differential equation
\begin{equation}
- \frac{d^2 \psi(y)}{d y^2} + \frac{\Omega^2}{4} \psi(y) =
\lambda \psi(y)
\end{equation}
which can be related to the regular Sturm--Liouville
eigenvalue problem
via $p(y) = 1 > 0$; $w(y) = 1 > 0$ and $q(y) = \Omega^2/4$.

The boundary conditions given by (\ref{eq:13.40}) and
(\ref{eq:13.43}) can
be expressed as
\begin{eqnarray}
\frac{\Omega}{2} \cdot \psi(y=0) + \left(-1 \right) \cdot
\left. \frac{d
\psi}{dy} \right|_{y=0} 
&=&0 \; ,  \\
1 \cdot \psi(y=1) + 0 \cdot \left. \frac{d \psi}{dy}
\right|_{y=1} 
&=&0 
\end{eqnarray}
in agreement with (\ref{sl02}) and (\ref{sl03}).

The up--to--now unknown separation constant $\lambda$ has a
spectrum of
real positive eigenvalues which can be calculated using
(\ref{sl09}) from
\begin{equation}
\lambda_n = \int_0^1 \left[ \left( \frac{d\psi_n(y)}{dy}
\right)^2 
+ \frac{\Omega^2}{4} \psi_n(y)^2\right] dx \; 
- \; \left| \psi_n(y) \frac{d\psi_n(y)}{dy} \right|_0^1 
\end{equation}
taking into account normalized orthogonal eigenfunction
(\ref{sl10})
\begin{equation}
\int_0^1 \psi_n(y) \psi_m(y) dy = \delta_{mn} \; .
\end{equation}
%%%%%%%%%%%%%%%%%%%%%%%%%%%%%%%%%%%%%%%%%%%
\section{Conclusions}
The presented paper shows the analytical method how to solve
drift--diffusion initial--boundary--value problem for the
case of reflecting and absorbing
boundaries~\cite{phr2,phr1,lin1,lin2,red}. On the basis of
Sturm--Liouville theory, the set of eigenvalues with
corresponding eigenfunctions has been found. Here we have
paid our attention to wave number calculations from
transcendental equations. The equations have been solved
numerically by Newton method. The main problem which has
been
solved was the dependence of obtained
results on drift value, i.~e. different cases of
control parameter $\Omega<-2$, $\Omega = -2$ and $\Omega>
-2$.
First case of $\Omega<-2$ corresponds to the situation when
it
is difficult and probably impossible, with significant
small
probability and for long times only, to leave the interval
due
to the large negative value of drift. 
The case of $\Omega = -2$ has been considered as limit
case and the corresponding solution has been found. The
opposite case of $\Omega>-2$ shows the usual situation 
when the system reaches the right border relatively fast. 
As application, the first passage time distribution as well
as the cumulative probability have been calculated. The
case of large positive values of $\Omega$ has been
investigated in detail and has been obtained
as approximation.
%%%%%%%%%%%%%%%%%%%%%%%%%%%%%%%%%%%%%%%%%%%%%
\section*{Acknowledgement}
One of us (J.~Hinkel) gratefully acknowledge support by
Graduiertenkolleg~567 \textit{Strongly Correlated
Many-Particle Systems}. The
authors are indebted to R.~K\"uhne (Berlin), J.~Kaupu\v{z}s
(Riga), I. Lubashevsky (Moscow), E.~Shchukin and
J.~Ro{\ss}mann (Rostock) for
fruitful discussions.
%%%%%%%%%%%%%%%%%%%%%%%%%%%%%%%%%%%%%%%%%%%%%

\end{document}